\documentclass[journal,twoside,web]{ieeecolor}
\usepackage{tmi}
\usepackage{cite}
\usepackage{amsmath,amssymb,amsfonts}
\usepackage{algorithmic}
\usepackage{graphicx}
\usepackage{textcomp}
\def\BibTeX{{\rm B\kern-.05em{\sc i\kern-.025em b}\kern-.08em
    T\kern-.1667em\lower.7ex\hbox{E}\kern-.125emX}}
\usepackage{footnote}
\usepackage{booktabs}
\usepackage{algorithm}
\usepackage{threeparttable}
\usepackage{multirow}
\usepackage{indentfirst}
\usepackage[misc]{ifsym}
\newtheorem{theorem}{Theorem}
\usepackage
[bookmarks=true, 
hyperfootnotes=false,
]
{hyperref}

\usepackage{graphics}

\usepackage{xcolor}

\usepackage[colorinlistoftodos, textwidth=10mm]{todonotes}
\usepackage{marginnote}

\begin{document}
\title{Brain Age Estimation From MRI Using Cascade Networks with Ranking Loss}

\author{{Jian Cheng, Ziyang Liu, Hao Guan, Zhenzhou Wu, Haogang Zhu, Jiyang Jiang, Wei Wen, Dacheng Tao,~\IEEEmembership{Fellow,~IEEE} and Tao Liu}
\thanks{Manuscript received XX XX, 2020. 
This research received support from the National Natural Science Foundation of China (Grant No. 61971017 and 81871434), 
the National Key Research and Development Program of China (Grant No. 2019YFC0118602) and Beijing Natural Science Foundation (Grant No. Z200016). 
\emph{Corresponding author: Tao Liu.}}
\thanks{Jian Cheng and Ziyang Liu contributed equally to this work. }
\thanks{Jian Cheng and Haogang Zhu are with the Beijing Advanced Innovation Center for Big Data-Based Precision Medicine, Beijing, China, and also with School of Computer Science and Engineering, Beihang University, Beijing, China (e-mail: jian\_cheng@buaa.edu.cn; haogangzhu@buaa.edu.cn).}
\thanks{Ziyang Liu is with the Beijing Advanced Innovation Center for Biomedical Engineering, the School of Biological Science and Medical Engineering, Beihang University, China (e-mail: liuziyang1106@buaa.edu.cn).}
\thanks{Hao Guan is with UBTech Sydney Artificial Intelligence Institute, School of Computer Science, FEIT, University of Sydney, Darlington, NSW, Australia (e-mail: hgua8781@uni.sydney.edu.au).}%
\thanks{Dacheng Tao is with JD Explore Academy at JD.com, China (e-mail: dacheng.tao@gmail.com).}%
\thanks{Zhenzhou Wu is with the BioMind Technology Center, Beijing TianTan Hospital, Beijing, China (e-mail: joe.wu@biomind.ai).}
\thanks{Jiyang Jiang and Wei Wen are with Centre for Healthy Brain Ageing, School of Psychiatry, University of New South Wales, Sydney, NSW, Australia (e-mail: jiyang.jiang@unsw.edu.au; w.wen@unsw.edu.au).}
\thanks{*Tao Liu is with the Beijing Advanced Innovation Center for Biomedical Engineering, the School of Biological Science and Medical Engineering, Beihang University, China. He is also with the Beijing Advanced Innovation Center for Big Data-Based Precision Medicine, Beijing, China (e-mail: tao.liu@buaa.edu.cn).}
}%

\markboth{IEEE TRANSACTIONS ON Medical Imaging,~Vol.~xx, No.~xx,~2020}{Cheng \MakeLowercase{\textit{et al.}}: Brain age estimation using cascade networks with ranking loss}
\maketitle

\begin{abstract}

Chronological age of healthy people is able to be predicted accurately using deep neural networks from neuroimaging data, 
and the predicted brain age could serve as a biomarker for detecting aging-related diseases. 
In this paper, a novel 3D convolutional network, called two-stage-age-network (TSAN), is proposed to estimate brain age from T1-weighted MRI data. 
Compared with existing methods, TSAN has the following improvements. 
First, TSAN uses a two-stage cascade network architecture, where the first-stage network estimates a rough brain age, 
then the second-stage network estimates the brain age more accurately from the discretized brain age by the first-stage network. 
Second, to our knowledge, TSAN is the first work to apply novel ranking losses in brain age estimation, together with the traditional mean square error (MSE) loss. 
Third, densely connected paths are used to combine feature maps with different scales. 
The experiments with $6586$ MRIs showed that TSAN could provide accurate brain age estimation, 
yielding mean absolute error (MAE) of $2.428$ and Pearson's correlation coefficient (PCC) of $0.985$, between the estimated and chronological ages. 
Furthermore, using the brain age gap between brain age and chronological age as a biomarker, 
Alzheimer's disease (AD) and Mild Cognitive Impairment (MCI) can be distinguished from healthy control (HC) subjects by support vector machine (SVM). 
Classification AUC in AD/HC and MCI/HC was $0.904$ and $0.823$, respectively. 
It showed that brain age gap is an effective biomarker associated with risk of dementia, and has potential for early-stage dementia risk screening. 
The codes and trained models have been released on GitHub: \url{https://github.com/Milan-BUAA/TSAN-brain-age-estimation}.

\end{abstract}

\IEEEpeerreviewmaketitle
\section{Introduction}
\label{sec:introduction}
\IEEEPARstart{W}{ith} increasing age, human body grows old and eventually stops functioning. 
The aging process of human brain is known to be biologically complex~\cite{lopez:cell2013:aging}, 
because human brains accumulate damage at different rates, 
and sometimes the aging process is even accelerated and hijacked by underlying pathology (e.g., neurodegenerative disease)\cite{bijsterbosch2019old}. 
Therefore, it is crucial to find biomarkers which can predict who has a higher risk of age-related deterioration, 
how the functional decline will progress, and which treatment is the most effective one.

Aging results in significant changes in brain structure and function. 
Manifold pathological changes begin to develop years or decades before the onset of cognitive decline~\cite{jack2010hypothetical}, 
such as abnormal changes in brain structures already at mild cognitive impairment (MCI) stage~\cite{driscoll2009longitudinal,spulber2010whole}. 
Additionally, atrophic regions detected in Alzheimer's disease (AD) patients were found to largely overlap with those regions showing in a normal age-related decline in health control subjects~\cite{dukart2011age}.
It was reported that, by 2050, $1$ in $85$ persons worldwide will be affected by AD, the most common form of dementia~\cite{brookmeyer2007forecasting}. 
Dementia often accompanies the abnormal aging of the brain~\cite{franke:NI2010:age,Gaser2013covert,ashburner2003}.
Brain morphometric pattern analysis has been increasingly investigated to identify age-related imaging biomarkers from structural magnetic resonance imaging (MRI)~\cite{liu:miccai2017}. 
Compared with other biomedical image modalities, MRI is a non-invasive means of potentially identifying abnormal structural brain changes in a more sensitive manner~\cite{liu:miccai2017}.

Chronological ages of healthy people can be accurately estimated from structural or functional brain neuroimaging data~\cite{franke:NI2010:age,dosenbach:science2010:age,cole:NI2017,cole:NI2020,cole:tn2017:age,cole2018brain,Jonsson2019,Peng2019}. 
The estimated age is normally called ''\emph{brain age}''. 
Brain age has potential to serve as a biomarker to explore brain aging process, cognitive aging, and age-related diseases~\cite{franke:NI2010:age,dosenbach:science2010:age,cole:tn2017:age}.
In general, the brain of healthy subjects follows a normal aging trajectory.
Thus, an effective approach to brain age estimation is to use machine learning methods on T1-weighted MR images of healthy subjects. 
Traditional machine learning methods of classification or regression have been proposed to estimate brain age. 
Classification based age estimation is to classify MRIs into different discrete brain age ranges~\cite{su:2011age,pardakhti:2017}, 
which normally has large errors due to false classification or a large age range. 
Various regression methods have been proposed to predicate brain age using different features extracted from MRIs, 
including linear regression with regularizations~\cite{ning:2018}, relevance vector machines~\cite{franke:NI2010:age,madan:2018predicting,luders:NI2016:age,lancaster:2018:age}, Gaussian process~\cite{cole:2015prediction}, hidden Markov models~\cite{wang:2011:age}, random forest~\cite{Liem2017}, etc.  

Recently, deep neural network based machine learning methods have been increasingly used in neuroimaging and brain age estimation~\cite{cole:NI2017,cole2018brain,Jonsson2019,franke:NI2012:age,dinsdale:NI2020}. 
By learning the correspondence between patterns in T1 images and age labels, 
deep learning algorithms can serve as high-dimensional regression models, which predict chronological ages from 3D T1 images. 
Existing methods mainly use different architectures of 3D convolutional neural networks (CNN). 
Cole \emph{et al.}~\cite{cole:NI2017,cole:NI2020} trained a VGG network~\cite{simonyan:2014:vgg} and obtained promising brain age estimation results with the Mean Absolute Error (MAE) of $4.65$ year. 
A ResNet design was used for brain age estimation in~\cite{Jonsson2019}.  
Peng \emph{et al.}~\cite{Peng2019} proposed Simple Fully Convolutional Network (SFCN) and won Predictive Analytics Competition 2019 (PAC-2019) with $\text{MAE}=2.90$. 
MAEs of predicated brain ages are generally around 4-5 years, when chronological ages are 18-90 years~\cite{cole:NI2020}. 
A larger training dataset and a smaller chronological age range in the test data normally result in a better estimation with smaller MAE~\cite{cole:NI2020}. 
However, CNNs in~\cite{cole:NI2017,cole:NI2020,Jonsson2019,Peng2019} have some limitations.  
1) They do not explore feature maps from different scales, compared with more advanced DenseNet~\cite{huang:cvpr2017:densenet}; 
2) They only consider the MAE loss between estimated and true ages of individual samples, 
without using any ranking loss for a set of samples. 
However, a good estimation means not only low MAE loss between individual samples, but also low ranking loss for a batch of samples.
Moreover, PAC-2019 considered Spearman's rank correlation coefficient (SRCC) as a criteria in evaluation, which has not yet been considered in training. 
3) Except~\cite{Jonsson2019}, they (both CNN in~\cite{cole:NI2017} and SFCN in~\cite{Peng2019}) mostly do not consider biological sex labels of MRIs, 
while males and females have different brain structures and ages differently~\cite{franke:NI2010:age,dosenbach:science2010:age,cole:tn2017:age}. 

The difference between the predicted age and the chronological age is called ''\emph{brain age gap}'' or ''\emph{brain age difference}''\cite{franke:NI2010:age,Smith2019,Gaser2013covert,franke2012longitudinal}. 
The positive brain age gap is assumed to reflect accelerated brain aging and the negative brain age gap means decelerated or healthy brain aging. 
Measuring brain age gap in patients is helpful in quantifying heterogeneity of diseases and improving disease risk screening. 
Brain age gap has been proved to be associated with a range of biological and cognitive variables~\cite{Cole2019bodyage,de2020cardioRisk,Smith2019,schnack2016accelerated,Nenadic2017,cole:tn2017:age,Cole2017a,cole2018brain,Franke2019}. 
The brain age concept has recently been in investigating neurodegenrative disorders to enable the early diagnosis of MCI and AD~\cite{Gaser2013covert,Liem2017,Beheshti2020,Wang2019}.

An observed bias frequently occurs in brain age estimation: 
brain age is often overvalued in younger subjects and underestimated in older subjects, while brain age for participants with an age closer to the mean age are predicted more accurately~\cite{Smith2019,Liang2019,de2020commentary}. 
It is possible to apply a statistical bias correction to age estimation or brain age gap estimate~\cite{cole2018brain,beheshti2019bias,Liang2019}. 

In this work, a novel 3D CNN architecture with ranking losses is proposed for brain age estimation. 
This paper has the following contributions:
\begin{itemize}
  \item A two-stage cascade network architecture, called two-stage-age-network (TSAN), is proposed.  
    Specifically, the first-stage network estimates a rough brain age, 
    and the second-stage network estimates a more accurate brain age based on the discretized brain age estimated by the first-stage network. 
    Biological sex label is used as an input for the above two networks, considering sex is a known scanning metadata.  
  \item Inspired by DenseNet~\cite{huang:cvpr2017:densenet}, 
    A novel scaled dense (ScaledDense) network architecture is used in both stages to combine feature maps with different scales by using densely connected paths. 
  \item Besides the normally used mean square error (MSE) loss, two ranking losses are proposed for regularizing the training process. 
    For two samples, the first ranking loss is MSE between the chronological and estimated age differences. 
    For a set of samples, the second ranking loss is defined by using Spearman's rank correlation coefficient (SRCC) of the chronological and estimated ages. 
  \item Based on a linear regression model, bias correction is applied to brain age estimation. 
    We demonstrate that estimated age by TSAN has lower MSE and a better classification result after bias correction. 
  \item In order to demonstrate applications of brain age estimation, 
    we use the brain age gap as the only input variable to classify healthy control subjects, MCI and AD patients by support vector machine (SVM)~\cite{bishop:PRML2006}.
\end{itemize} 

A preliminary version of this work was presented at a conference~\cite{liu:2020:age}. 
The work is significantly extended in this paper with more methodological details, experiments, validations, and discussions. 
We highlight some improvement of the materials in this paper over the conference paper. 
1) The proposed TSAN was compared with both CNN in~\cite{cole:NI2017} and SFCN in~\cite{Peng2019}. 
2) More experiments on model comparisons were performed with different losses and hyper-parameters. 
The experiments showed that the proposed ranking losses not only work for the proposed TSAN, but also improves existing methods (CNN and SFCN). 
3) We considered bias correction in the brain age estimation, which provides more accurate brain age and brain age gap.
4) Brain age gap was used in SVM to classify healthy control subjects, MCI and AD patients. 
The proposed ensemble TSAN after bias correction generally yields the best classification performance.
We have released our codes and trained models on GitHub\footnote{\url{https://github.com/Milan-BUAA/TSAN-brain-age-estimation}}.

\begin{figure}[!t]
  \centering
  \includegraphics[width=\linewidth]{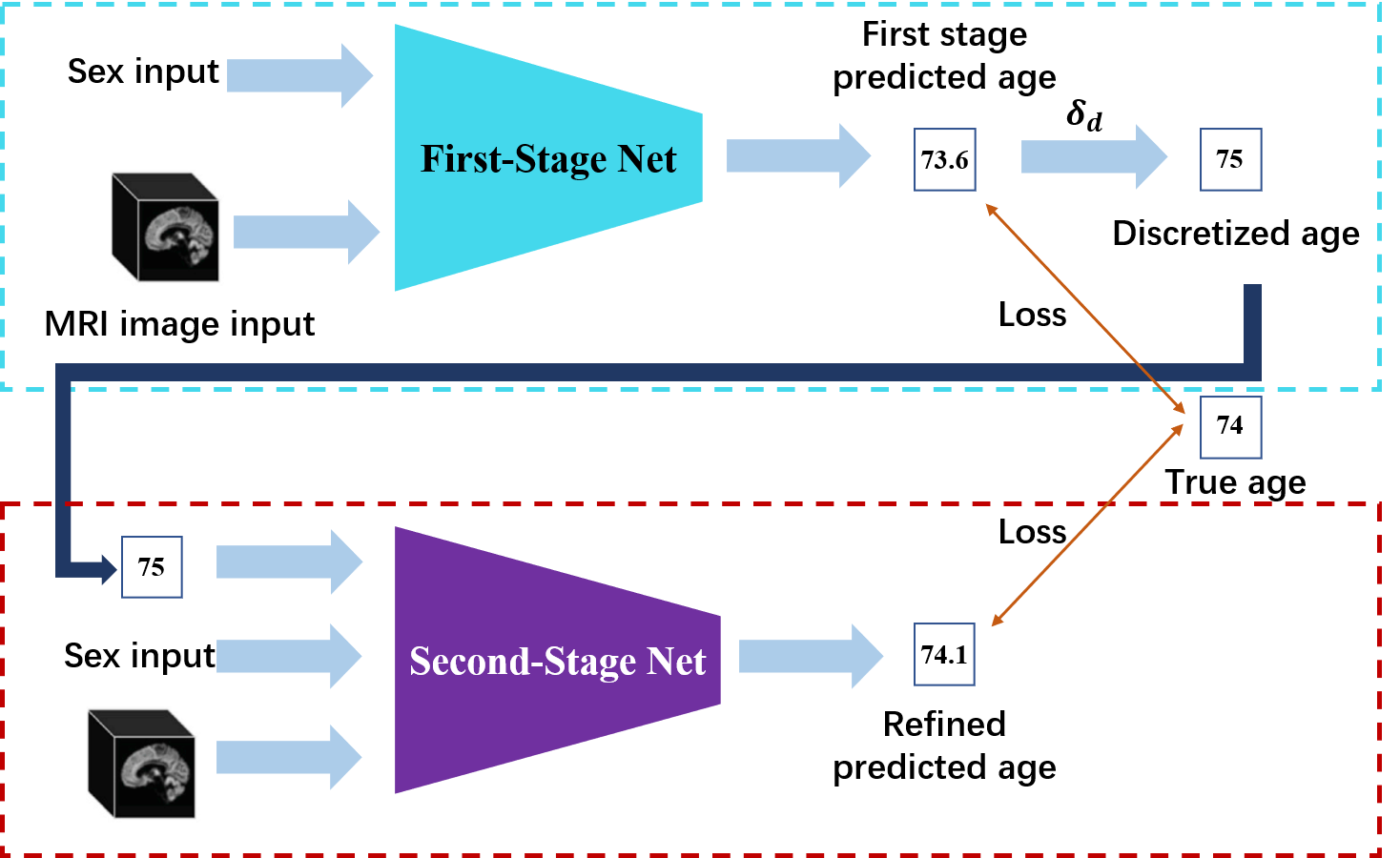}
  \caption{\label{fig:TSAN}\textbf{Two-Stage-Age-Network (TSAN).}   
  The first-stage network roughly estimates the brain age.  
  Then, the second-stage network estimates a more accurate age based on the discretized brain age estimated by the first-stage network. 
  Both networks use the sex label and MRI data as inputs.
  }
\end{figure}

\section{Method}

\subsection{Brain Age Estimation Framework}

\subsubsection{Two-Stage-Age-Network (TSAN)} 
\label{sec:TSAN}

\begin{figure*}[!t]
  \centering
  \includegraphics[width=0.7\linewidth]{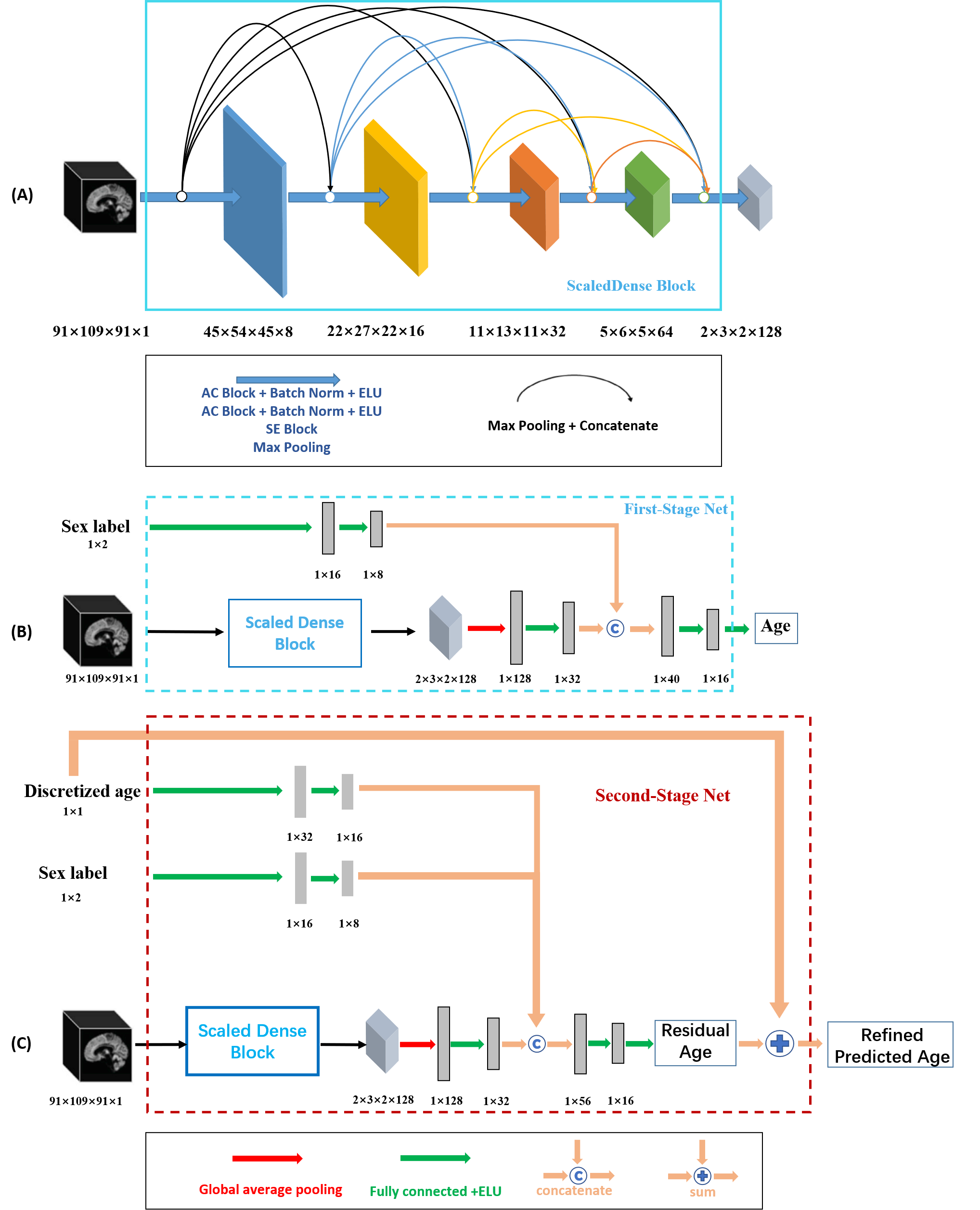}
  \caption{\label{fig:TSAN_network}\textbf{Detailed network architectures in TSAN.} 
  \textbf{(A)}: The ScaledDense block, where feature maps from different scales are combined by using pooling and concatenation. 
  Each layer of the ScaledDense block is designed by using Asymmetric Convolution (AC) blocks~\cite{ding:CVPR2019:acnet}, 
  batch norm~\cite{ioffe2015batch}, Exponential Linear Unit (Elu), Squeeze-and-Excitation (SE) block~\cite{hu:CVPR2018:se} and max pooling. 
  \textbf{(B)}: The first-stage network to roughly estimate the brain age from a sex label and an MRI. 
  \textbf{(C)}: The second-stage network to estimate a more accurate brain age from the sex label, MRI, and the discretized brain age by the first-stage network.
  }
\end{figure*}

Fig.~\ref{fig:TSAN} is an illustration map of TSAN. 
The first-stage network roughly estimates the brain age from a sex label and an MRI. 
Then, the second-stage network estimates a more accurate brain age from the sex label, 
MRI, and the discretized brain age estimated by the first-stage network. 
The cascade architecture of algorithms has been increasingly used in machine learning and computer vision~\cite{sun:CVPR2013:cascade}. 
The main contribution of the cascade architecture in TSAN is that 
the \emph{discretized} brain age (not the brain age itself) estimated by the first-stage network is used as an input of the second-stage network.  
Considering existing works~\cite{cole2018brain,beheshti2019bias,Liang2019} on brain age estimation have good estimations with $\text{MAE}<5$, 
It is easier for the first-stage network to predict a discretized brain age (i.e., which age range the ground truth age belongs to), 
compared with predicting a continuous brain age. 

TSAN uses $\delta_d$ to discretize the estimated brain age in $\mathbb{R}^1$. 
See Fig.~\ref{fig:TSAN}. 
Let $\hat{y}$ be the estimated brain age, and $y$ be the true chronological age. 
Then, the discretized brain age $D(\hat{y})$ is defined as  
\begin{equation}
  D(\hat{y}) = \left\{
    \begin{array}{lr}
      \text{Round}(\frac{\hat{y}}{\delta_d}) \cdot \delta_d, &  \text{if}\ \delta_d > 0 \\
      \hat{y}, &  \text{if}\ \delta_d = 0
    \end{array} 
    \right. \label{eq:dist}
\end{equation}%
where $\delta_d$ is a non-negative parameter for tuning the discretization degree, and $\text{Round}(\cdot)$ denotes the round operator. 
Based on Eq.~\ref{eq:dist}, if we set $\delta_d=0$, then there is no discretization, i.e., $D(\hat{y})=\hat{y}$. 
If $\delta_d=5$, then $D(73.6)=75$, as demonstrated in Fig.~\ref{fig:TSAN}. 
Note that the first-stage network together with a positive $\delta_d$ could be seen as a multi-class classify 
which determines which age range the ground truth age $y$ belongs to.

Theorem~\ref{thm:err} demonstrates that, if $\delta_d>|y-\hat{y}|$, 
then the discretized brain age $D(\hat{y})$ can be a good approximation of both the chronological age $y$ and its discretized version $D(y)$. 
Considering the discretization is in $\mathbb{R}^1$, the proof of theorem~\ref{thm:err} is trivial. 

\begin{theorem}[\textbf{Discretization Approximation}]\label{thm:err}
  Let $\delta_d$ be the discretization parameter in Eq.~\eqref{eq:dist}, and $\epsilon=|\hat{y}-y|$ be the absolute error of brain age estimation. 
  If we set $\delta_d \geq \epsilon$, then  
  1) $|D(y)-D(\hat{y})|=0\ \text{or}\ \delta_d$; 
  2) $|D(\hat{y})-y| \leq \frac{1}{2}\delta_d + \epsilon \leq \frac{3}{2} \delta_d$. 
\end{theorem}

Theorem~\ref{thm:err} implies that 
we could design the second-stage network to estimate only a residual age $D(\hat{y})-y$ based the on discretized brain age $D(\hat{y})$ by the first-stage network, 
and the residual is proved to be bounded because of $|D(\hat{y})-y| \leq \frac{3}{2}\delta_d$. 

Fig.~\ref{fig:TSAN_network} (B) and (C) demonstrate network architectures in these two stages.  
In both stages, the input sex label, 
which is represented as a 2-dimensional one-hot vector, is concatenated with the feature vectors after global average pooling by Fully Connected (FC) layers. 
The discretized brain age by the first-stage network is used as an input of the second-stage network, 
and the second-stage network actually learns a residual age. 
The final estimated brain age is obtained by summing the residual age and the discretized brain age by the first-stage network. 

\subsubsection{ScaledDense Block}
\label{sec:scaleddensenet}

We propose a novel convolution block, called ScaledDense, as shown in Fig.~\ref{fig:TSAN_network} (A). 
The ScaledDense block is used in both networks of the two stages. 
See Fig.~\ref{fig:TSAN_network} (B) and (C). 
The ScaledDense block is inspired by DenseNet~\cite{huang:cvpr2017:densenet}. 
The densely connected paths in DenseNet~\cite{huang:cvpr2017:densenet} combine feature maps with the same scale size from preceding layers by concatenation. 
Specifically, in the DenseNet block, the $l$-th layer receives the feature-maps of all preceding layers with the same size, 
i.e., $x_l=H_l([x_0, x_1,\cdots, x_{l-1} ])$, where $x_i$ is the output feature map of the $i$-th layer, and $H_l$ is the non-linear transformation of the $l$-th layer. 
Note that feature maps $\{x_0, x_1,\cdots, x_{l-1}\}$ should have the same size for concatenation.  
While in the ScaledDense block, the densely connected paths concatenate feature maps with different scale sizes from preceding layers by resizing (pooling, upsampling, etc.) and concatenation, 
i.e., 
\begin{equation}
 x_l=H_l([S_0^l(x_0), S_1^l(x_1),\cdots, S_{l-1}^l(x_{l-1}) ]), 
\end{equation}
where $S_i^l$ is the resizing operator from $x_i$ to $x_l$. 
In this way, feature maps  $\{x_0, x_1,\cdots, x_{l-1}\}$ could have different scale sizes. 
The resizing operator $S_i^l$ could be pooling, upsampling, transpose convolution, etc, depending on the sizes of $x_i$ and $x_l$. 
In Fig.~\ref{fig:TSAN_network} (A), the resizing operator is pooling because the feature maps of the preceding layers have larger sizes than the feature map $x_l$ in the $l$-th layer.
Similarly with DenseNet~\cite{huang:cvpr2017:densenet}, every layer in the ScaledDense block has direct connections to the loss function and the input signal, 
which results in better parameter efficiency and an implicit deep supervision.  

The ScaledDense block could use different non-linear transformations, similarly with DenseNet~\cite{huang:cvpr2017:densenet}. 
In this paper, each layer of the ScaledDense block is designed by two Asymmetric Convolution (AC) blocks~\cite{ding:CVPR2019:acnet}, 
batch norm~\cite{ioffe2015batch}, Exponential Linear Unit (Elu) activation function, Squeeze-and-Excitation (SE) block~\cite{hu:CVPR2018:se} and max pooling. 
The original AC block in~\cite{ding:CVPR2019:acnet} was designed for natural images with three 2D Conv (i.e., $d\times d$, $d\times 1$, and $1\times d$). 
Thus, the AC block for MRIs is designed with four 3D Conv (i.e., $d\times d\times d$, $d\times 1\times 1$, $1\times d \times 1$, and $1\times 1 \times d$). 
We found in experiments that this setting with two AC blocks ($d=3$) and one SE block is better than two traditional convolutional blocks for this brain age estimation task. 

Since the chronological age is positive, 
the activation function of the last FC layer in the first-stage network in Fig.~\ref{fig:TSAN_network} (B) is set as the Rectified Linear Unit (ReLu) function, in order to output positive brain age.
Since the residual age learned by the second-stage network is ideally in $(-\infty, \infty)$, 
the activation function of the last FC layer before the residual age in Fig.~\ref{fig:TSAN_network} (C) is set as the identity mapping. 
In this way, even if the brain age estimated by the first-stage network is far from the ground truth age, 
the second-stage network could still estimate a more accurate brain age with a residual age probably larger than $\delta_d$.

\subsubsection{Loss Functions} 
\label{sec:loss}

The mean absolute error (MAE) loss in Eq.~\eqref{eq:mae} and mean square error (MSE) loss in Eq.~\eqref{eq:mse} 
are two standard losses for training deep networks in brain age estimation, 
where $N$ is the batch size. 
\begin{equation}\label{eq:mae}
  \mathcal{L}_{\text{MAE}} = \frac{1}{N}\sum_i |\hat{y}_i-y_i|.
\end{equation}
\begin{equation}\label{eq:mse} 
  \mathcal{L}_{\text{MSE}}=\frac{1}{N}\sum_i (\hat{y}_i-y_i)^2. 
\end{equation}
MAE is also used as a standard evaluation metric for results of brain age estimation~\cite{cole:NI2017}. 
In this paper, we use MSE to train TSAN, 
because MSE penalizes more for large errors when $\text{MAE}>1$, 
and MAE by existing methods is normally larger than $1$~\cite{franke:NI2010:age,dosenbach:science2010:age,cole:NI2017,cole:NI2020,cole:tn2017:age,cole2018brain,Jonsson2019,Peng2019}. 
Both MAE in Eq.~\eqref{eq:mae} and MSE in Eq.~\eqref{eq:mse} are defined for individual samples. 
However, relationships among two or more samples are also important, including difference between two ages, and the ranking order of ages.

For two samples with ages $y_i$ and $y_j$, 
the age difference loss in Eq.~\eqref{eq:age_difference} is proposed as the MSE between the estimated brain age difference $\hat{y}_i -\hat{y}_j$ and true age difference $y_i-y_j$, 
where $N_p$ denotes the number of paired samples $(i,j)$. 
\begin{equation}\label{eq:age_difference}
  \mathcal{L}_{\text{d}} = \frac{1}{N_p} \sum_{(i,j)} ( (\hat{y}_i - \hat{y}_j) - (y_i-y_j) )^2, 
\end{equation}%
The age difference loss $\mathcal{L}_{\text{d}}$ can be seen as a ranking loss for two samples, because the sign of the age difference indicates the order of two ages. 
$\mathcal{L}_{\text{d}}$ considers both the magnitude and sign of the age difference.

In statistics, Spearman's rank correlation coefficient\footnote{\href{https://en.wikipedia.org/wiki/Spearman\%27s\_rank\_correlation\_coefficient}{https://en.wikipedia.org/wiki/Spearman's\_rank\_correlation\_coefficient}} 
(SRCC) is a non-parametric measure of rank correlation. 
  For $N$ samples, SRCC is defined as the Pearson's correlation coefficient (PCC) between the rank values of those two variables: 
\begin{equation}\label{eq:SRCC_1}
  r_r = \frac{\text{cov}(\text{Rank}(\hat{y}), \text{Rank}(y))}{\sigma_{\text{Rank}(\hat{y})} \sigma_{\text{Rank}(y)}}, 
\end{equation}%
  where $\text{Rank}(\cdot)$ is the rank operator, 
  $\text{cov}(\text{Rank}(\hat{y}), \text{Rank}(y))$ is the covariance of the two rank variables, 
  and $\sigma_{\text{Rank}(\hat{y})}$ and $\sigma_{\text{Rank}(y)}$ are the standard deviations of the two rank variables. 
  If there is no identical values in both $\{\hat{y}_i\}$ and $\{y_i\}$, 
  then the above definition is equivalent to 
\begin{equation}\label{eq:SRCC_2}
  r_r = 1- \frac{6 \sum_i (\text{Rank}(\hat{y}_i) - \text{Rank}(y_i))^2 } {N(N^2-1)}.  
\end{equation}%
If there are some identical values in $\{\hat{y}_i\}$ or $\{y_i\}$, 
  $\text{Rank}(\cdot)$ indicates the fractional ranking operator\footnote{\url{https://en.wikipedia.org/wiki/Ranking\#Fractional_ranking_.28.221_2.5_2.5_4.22_ranking.29}}, 
which means that identical values are each assigned fractional ranks equal to the average of their positions in the ascending order of the values. 
In this paper, we define a ranking loss based on SRCC:  
\begin{equation}\label{eq:loss_rank}
  \mathcal{L}_\text{r} = \sum_i (\text{Rank}(\hat{y}_i) - \text{Rank}(y_i))^2.  
\end{equation}%
Note that if there is no identical values in both $\{\hat{y}_i\}$ and $\{y_i\}$, 
  then $\mathcal{L}_\text{r}$ is propotional to $1-r_r$. 
  Otherwise, $\mathcal{L}_\text{r}$ is an approximation of $1-r_r$. 
  In both cases, if $\mathcal{L}_\text{r}=0$, then $\{\hat{y}_i\}$ and $\{y_i\}$ have the same rank and SRCC $r_r=1$.
The above ranking loss based on the rank operator $\text{Rank}(\cdot)$ is not practical to be optimized via gradient descent, 
because the rank operator $\text{Rank}(\cdot)$ is not differentiable.

SoDeep~\footnote{\url{https://github.com/technicolor-research/sodeep}}~\cite{engilberge:CVPR2019:sodeep} was proposed to approximate the sorting of arbitrary sets of scores. 
SoDeep trains a proxy differentiable network $R$ as an approximation of the rank operator $\text{Rank}(\cdot)$ based on synthetic ranking data~\cite{engilberge:CVPR2019:sodeep}. 
Following SoDeep in~\cite{engilberge:CVPR2019:sodeep}, the differentiable network $R$ is designed as an LSTM network [50], and is trained by solving
\begin{equation}\label{eq:rank}
  \min_R \sum_{i=1}^N (R(z_i) - \text{Rank}(z_i))^2,  
\end{equation}
where $\{z_i\}_{i=1}^N$ is a synthetic sequential data with $N$ values, 
$\text{Rank}(z_i)$ is the ground truth rank of $z_i$, 
and $R(z_i)$ is the approximated rank of $z_i$.
Experiments in~\cite{engilberge:CVPR2019:sodeep} showed that approximated rank $R(z_i)$ after training is very close to the ground truth rank $\text{Rank}(z_i)$, 
and SoDeep could be used for training various rank-based losses.
Then, the pre-trained proxy differentiable network $R$ is used to replace the rank operator in the loss function. 
Thus, with SoDeep, we could represent different sorting evaluation metrics as functions of this differentiable sorter $R$, 
hence making sorting metrics differentiable and suitable as training losses. 
After $\text{Rank}(\hat{y}_i)$ is replaced with $R(\hat{y}_i)$, 
the loss 
  $\mathcal{L}_{\text{r}}=\sum_i (R(\hat{y}_i) - R(y_i))^2$ 
is differentiable, 
and this differentiable proxy of $\mathcal{L}_{\text{r}}$ is used for training. 
Note that SoDeep with the proxy differentiable network 
does not change the network architectures in Fig.~\ref{fig:TSAN} and~\ref{fig:TSAN_network}, 
only making the ranking loss $\mathcal{L}_{\text{r}}$ differentiable for backpropagation. 

Combining above losses, the total loss function $\mathcal{L}$ is 
\begin{equation}\label{eq:loss}
  \mathcal{L} = \mathcal{L}_{\text{MSE}} + \lambda_1 \mathcal{L}_{\text{d}} + \lambda_2\mathcal{L}_{\text{r}}, 
\end{equation}%
where $\lambda_1$ and $\lambda_2$ are regularization parameters. 
It should be noted that the optimal estimation network after training satisfies $\mathcal{L}_{\text{MSE}}=0$, 
and meanwhile $\mathcal{L}_{\text{d}}=0$ and $\mathcal{L}_{\text{r}}=0$. 
Thus, the proposed two ranking losses $\mathcal{L}_{\text{d}}$ and $\mathcal{L}_{\text{r}}$ are designed for regularizing the training process.

\subsection{Bias correction}

Brain age is often overvalued in younger subjects and underestimated in older subjects, 
while the estimation MAE is close to zero when the chronological age is close to the mean age~\cite{Smith2019,Liang2019,de2020commentary}. 
We could apply a statistical bias correction to age estimation or brain age gap estimate~\cite{cole2018brain,beheshti2019bias,Liang2019}.
The bias correction method relies on fitting a linear regression model of brain age gap against chronological age in Eq.~\eqref{eq:offset} through the training set. 
\begin{equation}\label{eq:offset}
\Omega \approx \alpha \cdot y + \beta, 
\end{equation}%
where $y$ is the chronological age, $\alpha$ is the slope, $\beta$ is the intercept, $\Omega$ is the brain age gap which is used as an offset for bias correction, 
Then, with the pre-fitted $\alpha$ and $\beta$, 
the offset can be subtracted from the estimated brain age to achieve a bias-reduced brain age estimation for each test sample:
\begin{equation}\label{eq:bias_correct}
  \widetilde{y}_c = \hat{y} -  (\alpha \cdot y + \beta)
\end{equation}
where $\widetilde{y}_c$ is the corrected brain age and $\hat{y}$ represents the estimated brain age. 
It should be noted that $\alpha$ and $\beta$ are obtained by fitting the linear regression model using training data, 
and then they are applied to the test data for bias correction.

The above bias correction in Eq.~\eqref{eq:bias_correct} requires the chronological age $y$. 
Thus, it may not be appropriate for evaluating brain age estimation, because the chronological age of the test data is the estimation target. 
However, it may help following brain age related analysis based on brain age gap, for example dementia classification in Section~\ref{sec:classification}.

\subsection{Dementia classification}
\label{sec:classification}

Cognitive decline and neurodegenrative diseases often accompany the abnormal aging of brain.
Brain age gap has been proved to be able to reflect neurodegenrative disorders and enable the early diagnosis of MCI and AD~\cite{Gaser2013covert,Liem2017,Beheshti2020,Chung2018,Cole2019bodyage}.

Support vector machine (SVM)~\cite{bishop:PRML2006} is a supervised learning algorithm, based on statistical theory and the structural risk minimization principle. 
SVM is optimized to maximize the margin around the hyperplane that separates samples of different classes. 
It is widely used in medical image analysis including computer-aided diagnosis~\cite{Bron_2014,Bron_2015,Zeng_2018}. 

In order to investigate applications of brain age gap as a biomarker in dementia classification, 
we used brain age gap as the only feature in the SVM classifier to identify healthy controls subjects, MCI and AD patients. 

\section{Experiments}

\subsection{Datasets and Preprocessing}

In this paper, three public datasets were used in experiments, 
which are 
the Open Access Series of Imaging Studies (OASIS)~\cite{marcus:2010:OASIS}, 
the Alzheimer’s Disease Neuroimaging Initiative (ADNI-1)~\cite{jack:2008:ADNI}, 
and Predictive Analytics Competition 2019 (PAC-2019). 
These datasets have subjects with different age range. 
Please see Table~\ref{tab:data} for the detailed information of datasets, where $N_{\text{img}}$ is the number of images. 

\begin{figure}[!t]
  \centering
  \includegraphics[width=1\linewidth]{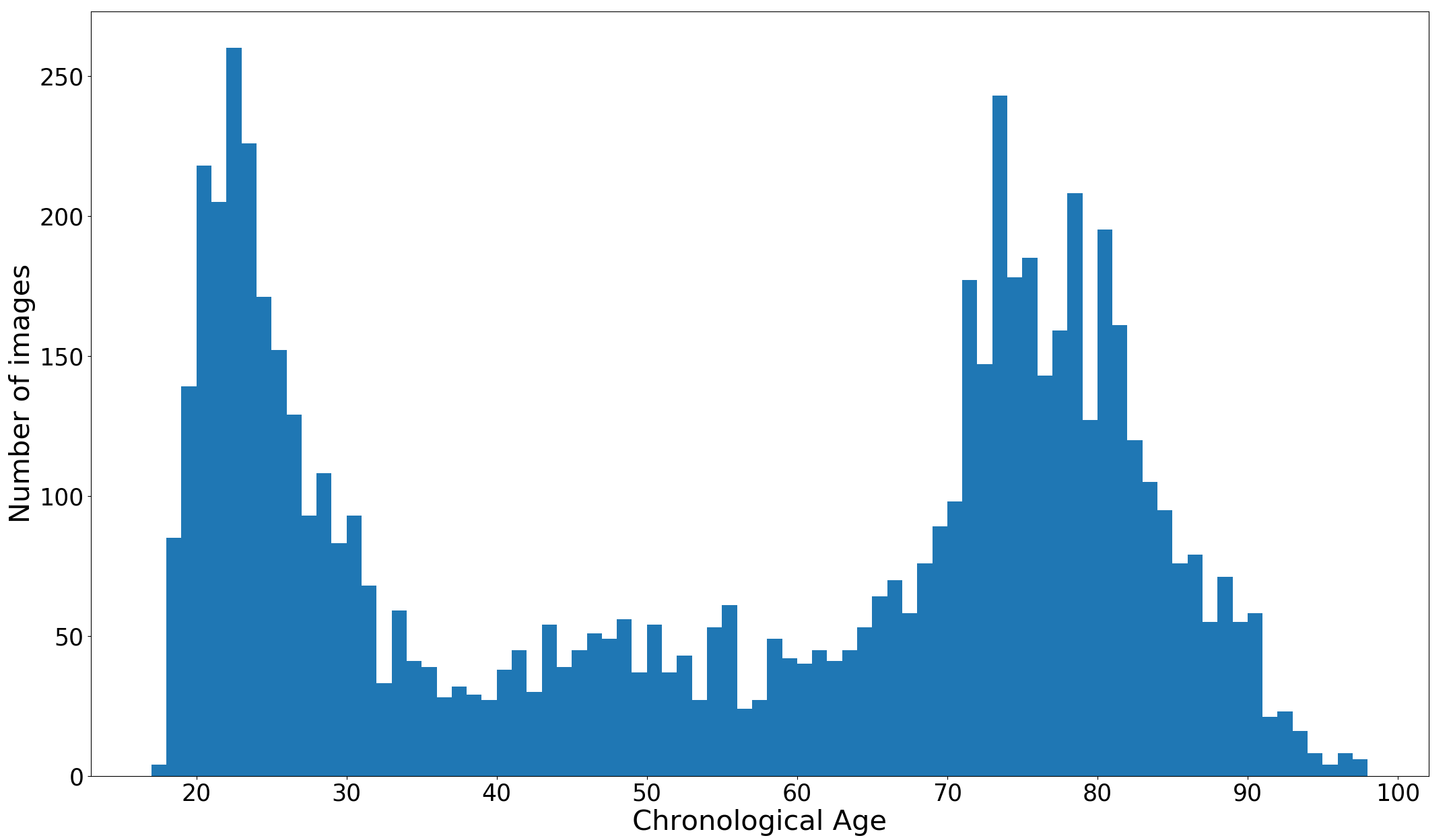}
  \caption{\label{fig:histogram_data}\textbf{The chronological age histogram of the combined dataset.}}
\end{figure}

For brain age estimation experiments, we combined these three data sets to make the age distribution as even as possible. 
Only healthy subjects among the combined data were used for training brain age estimation networks. 
Given the design of the ADNI and OASIS datasets, several follow-up scans were available for some subjects or multiple scans were performed in one acquisition, 
which introduced a natural type of data augmentation. 
The combined dataset has total $6586$ T1-weighted MRIs from healthy subjects. 
Fig.~\ref{fig:histogram_data} shows the chronological age histogram of the combined dataset. 

For dementia classification experiments using SVM, the ADNI-1 dataset was used. 
It is to investigate an application of the brain age gap. 
Table~\ref{tab:classfication data} shows the detailed information of the ADNI dataset in three groups, 
where $N_{\text{subject}}$ is the number of subjects. 
Note that there are $229$ HC subjects in ADNI-1, while we only consider $40$ HC subjects among them in ADNI-1 for dementia classification, 
because the other HC subjects were used in the training and validation datasets to train deep models in brain age estimation experiments. 

All MRIs in datasets were processed by using a standard preprocessing pipeline with FSL 5.10~\cite{woolrich2009bayesian,smith2004advances,jenkinson2012fsl}~\footnote{\url{https://fsl.fmrib.ox.ac.uk/fsl/fslwiki/FSL}}, 
including nonlinear registration~\cite{andersson2010non} to the standard MNI space, 
brain extraction~\cite{smith2002fast,jenkinson2005bet2}, 
and voxel value normalization (i.e., subtracting the mean and dividing the standard deviation of voxel values within the brain area).  
All MRIs after preprocessing have voxel size of $91\times 109\times 91$ with isotropic spatial resolution of $2$ $\text{mm}^3$. 

\begin{table*}[!t]
  \centering
  \caption{\label{tab:data}Demographic information of brain age estimation dataset. }
  \begin{tabular}{ccccc}
    \toprule
    Dataset & $N_{\text{img}}$&  Age Range&  Age Statistics (Mean $\pm$ STD) & Male / Female  \\ 
    \midrule
    ADNI & $1020$ & $60$ - $93$ & $77.4 \pm \ 5.08$ & $531$ / $489$  \\ 

    OASIS &$2926$ & $18$ - $98$ & $63.9 \pm 22.72$ & $1157$ / $1769$  \\ 

    PAC-2019 & $2640$ &$17$ - $90$ & $35.9 \pm 16.21$ & $1237$ / $1403$  \\ 

    Combined data & $6586$ & $17$ - $98$ & $54.7 \pm 24.44$ & $2925$ / $3661$ \\
    \bottomrule
  \end{tabular}
\end{table*}

\begin{table*}[!t]
  \centering
  \caption{\label{tab:classfication data}Demographic information of dementia classification dataset.}
  \begin{tabular}{cccccc}
    \toprule
    Group & $N_{\text{subject}}$&  Age Range& Mean age$\pm$STD& Mean brain age gap $\pm$ STD & MAE $\pm$ STD\\ 
    \midrule
    HC & $40$ & $70$ - $85$ & $76.1 \pm 4.39$ & $0.082 \pm 1.568$  & $1.067 \pm 1.150$\\ 

    MCI & $397$ & $55$ - $90$ & $74.8 \pm 7.40$ & $3.517 \pm 7.694$ & $5.004 \pm 4.013$ \\ 

    AD & $196$ & $55$ - $91$ & $75.7 \pm 7.68$ & $7.780 \pm 3.804$  & $9.471 \pm 4.124$ \\ 
    \bottomrule
  \end{tabular}
\end{table*}

\subsection{Experimental Settings}

To train and evaluate different learning models, 
we randomly split the combined dataset into 3 subsets: 
the training set ($70\%$, $4610$ MRIs), validation set ($15\%$, $988$ MRIs), and test set ($15\%$, $988$ MRIs). 
In ADNI-1 and OASIS datasets, there are several scans for one subject. 
The data was split under a subject level, 
meaning that all scans from the same subject will appear in just one of the training set, validation set, and test set.
The training-validation-test split was performed only once, and was used in all following experiments.

Three evaluation criteria were used for evaluating results of brain age estimation, 
including the mean absolute error (MAE), 
Pearson's correlation coefficient (PCC), and Spearman's rank correlation coefficient (SRCC). 
Note that PCC and SRCC can be calculated between the estimated brain age and the chronological age, 
which are denoted as PCC$_\text{age}$ and SRCC$_\text{age}$. 
On the other hand, PCC and SRCC can also be calculated between the estimated brain age gap and the chronological age, 
which are denoted as PCC$_\text{gap}$ and SRCC$_\text{gap}$. 
A good estimation is achieved when MAE, PCC$_\text{gap}$, and SRCC$_\text{gap}$ are close to $0$, and PCC$_\text{age}$ and SRCC$_\text{age}$ are close to $1$.

\subsection{Training of Brain Age Estimation Models}

For training TSAN, 
we first trained the first-stage network, 
and its output brain age was discretized in the step of $5$ years, i.e., $\delta_d=5$. 
See Fig.~\ref{fig:TSAN}. 
We set $\delta_d=5$ because $\text{MAE}<5$ for the first-stage network, 
and the second-stage network only estimates a bounded residual age based on Theorem~\ref{thm:err}. 
Then, we trained the second-stage network, while fixing the first-stage network. 

To train the first-stage network, the ADAM optimizer~\cite{Kingma:ICLR2015} was used for optimization, 
with the initial learning rate of $0.001$, decay $=10^{-6}$, $\beta_{1}= 0.9$,  $\beta_{2}= 0.999$, which are default in Pytorch. 
We set the batch size of $32$, the regularization weight $\lambda_1=\lambda_2=10$, based on the performance on the validation set. 
We used the He initialization strategy~\cite{He_2015_ICCV} for initialization, 
and considered an L2 norm weight regularization with the regularization weight of $5\times10^{-4}$. 
To compare different losses, 
we trained the first-stage network with $\mathcal{L}_{\text{MSE}}$ in Eq.~\eqref{eq:mse} and $\mathcal{L}$ in Eq.~\eqref{eq:loss}, respectively. 
Furthermore, in order to reduce overfitting, data augmentation was performed to input training MRIs with $50\%$ probability by a random spatial transformation, 
which combines a random 3D translation, a random rotation in the X-Y plane, and a random left-right flip of MR images. 
The translation distance was randomly chosen within $10$ voxels, and the rotation angle was randomly chosen between $-20^{\circ}$ and $20^{\circ}$. 

To train the second-stage network, a small initial learning rate $10^{-5}$ was used, considering the estimated brain age at the first stage is already close to the true chronological age. 
Other hyper-parameters were set the same as the first-stage network. 

Furthermore, we trained an ensemble TSAN model. 
TSAN with the total loss $\mathcal{L}$ was trained three times from different random initializations, 
and then the ensemble TSAN model is to ensemble these three models by averaging their estimated brain ages. 

We also implemented the CNN model in~\cite{cole:NI2017} and the SFCN model in~\cite{Peng2019} for comparison. 
The authors of~\cite{Peng2019} released a pre-trained SFCN model on github~\footnote{\url{https://github.com/ha-ha-ha-han/UKBiobank_deep_pretrain}}, 
which was trained by using the UK Biobank data. 
We tested SFCN on the combined dataset in Table~\ref{tab:data} in three ways.  
1) We directly applied the released pre-trained SFCN on the test set of the combined dataset. 
This model is called SFCN-pretrained. 
2) We trained SFCN from scratch with a random initialization by using the training set of the combined dataset. 
This model is called SFCN-train. 
3) We set the weight initialization of SFCN as the released pre-trained SFCN, and then trained SFCN by using the training set of the combined dataset. 
This model is called SFCN-transfer, because it transferred some knowledge from the UK Biobank dataset to brain age prediction on the combined dataset by using weight initialization. 
Since SFCN-pretrained was trained on the UK Biobank data, its performance on the combined data may not work well. 
SFCN-train and SFCN-transfer are assumed to have better results than SFCN-pretrained. 

All brain age estimation models were implemented in Pytorch.
The learning rate was decayed with a factor of $0.5$ when the training loss did not decrease within $5$ consecutive epochs. 
If the MAE in the validation set did not decrease within $20$ consecutive epochs, then the training was stopped. 

\subsection{Dementia Classification}

For dementia classification experiment, we considered three classification tasks including AD patients versus healthy control subjects (AD vs.\ HC), MCI patients versus HC (MCI vs.\ HC) and AD patients versus MCI patients (AD vs.\ MCI). 
For each task, we performed 5 fold nested stratified cross validation (CV) to tune model hyper-parameters with a grid search and evaluate performance of different brain age estimation models. 
The nested cross validation has an inner loop CV nested in an outer loop CV. 
The inner loop is a 5 fold stratified CV responsible to choose hyper-parameters with the best area under receiver operating characteristic (AUC). 
The outer loop is a another 5 fold CV for evaluating performance of the model with the chosen best hyper-parameters in the test set. 
We repeated the whole nested CV for $3000$ times. 
SVM classifier was chosen with radial basis function (RBF) kernel, and the search spaces for hyper-parameter $C$ and $\gamma$ in SVM are: 
$C\in \{1, 5\}\times\{10^{-2}, 10^{-1}, 1, 10^1, 10^2\}$, $\gamma \in \{1,5\}\times \{10^{-4},10^{-3},10^{-2},10^{-1},1\}$
The classification performance was evaluate by four metrics, including classification accuracy (ACC), sensitivity (SEN), specificity (SPE) and area under receiver operating characteristic (AUC). 
All classification tasks were performed using python and scikit-learn library~\cite{pedregosa2011scikit}.
and the class weights are automatically adjusted to be inversely proportional to class frequencies (i.e., the ``balanced'' mode).

\section{Result}

\subsection{Brian age estimation results without bias correction}

We trained all models (CNN~\cite{cole:NI2017}, SFCN~\cite{Peng2019}, and TSAN) by using different loss functions and hyper-parameters, 
and then the evaluation metrics (MAE, PCC$_\text{age}$, SRCC$_\text{age}$, PCC$_\text{gap}$, and SRCC$_\text{gap}$) were calculated by the trained models in the test dataset. 

We showed ablation experiments for choosing the hyper-parameters and loss function in TSAN in Table~\ref{tab:sex},~\ref{tab:delta}, and~\ref{tab:layer}.
Table~\ref{tab:sex} shows that considering sex labels in networks improves the brain age estimation results. 
  Different discretization parameter $\delta_d$ values were tested in Table~\ref{tab:delta}, where $\delta_d=5$ yields the best MAE for both the total loss and the MSE loss. 
  With the total loss, increasing $\delta_d$ from $5$ to $9$ could slightly improve PCC$_\text{age}$, SRCC$_\text{age}$, but obtain lower MAE. 
  Both Table~\ref{tab:sex} and Table~\ref{tab:delta} demonstrate that the total loss $\mathcal{L}$, 
  which incorporates both the MSE loss and ranking losses ($\mathcal{L}_{\text{d}}$ and $\mathcal{L}_{\text{r}}$), obtains lower MAE values than the MSE loss. 
  In the ScaledDense block in Fig.~\ref{fig:TSAN_network} (A), the number of layers $N_{layer}$ and number of channels in the initial layer $N_{ini}$ are two hyper-parameters, 
  and the number of channels in the $i$-layer is $2^i N_{ini}$. 
  Table~\ref{tab:layer} shows brain age estimation results with different $N_{layer}$ and $N_{ini}$, 
  where $N_{layer}=5$ and $N_{ini}=8$ yield the lowest MAE. 
  Thus, we set $\delta_d=5$, $N_{layer}=5$ and $N_{ini}=8$, and consider the total loss and sex labels in TSAN. 

\begin{figure*}[ht]
  \centering
  \includegraphics[width=1\linewidth]{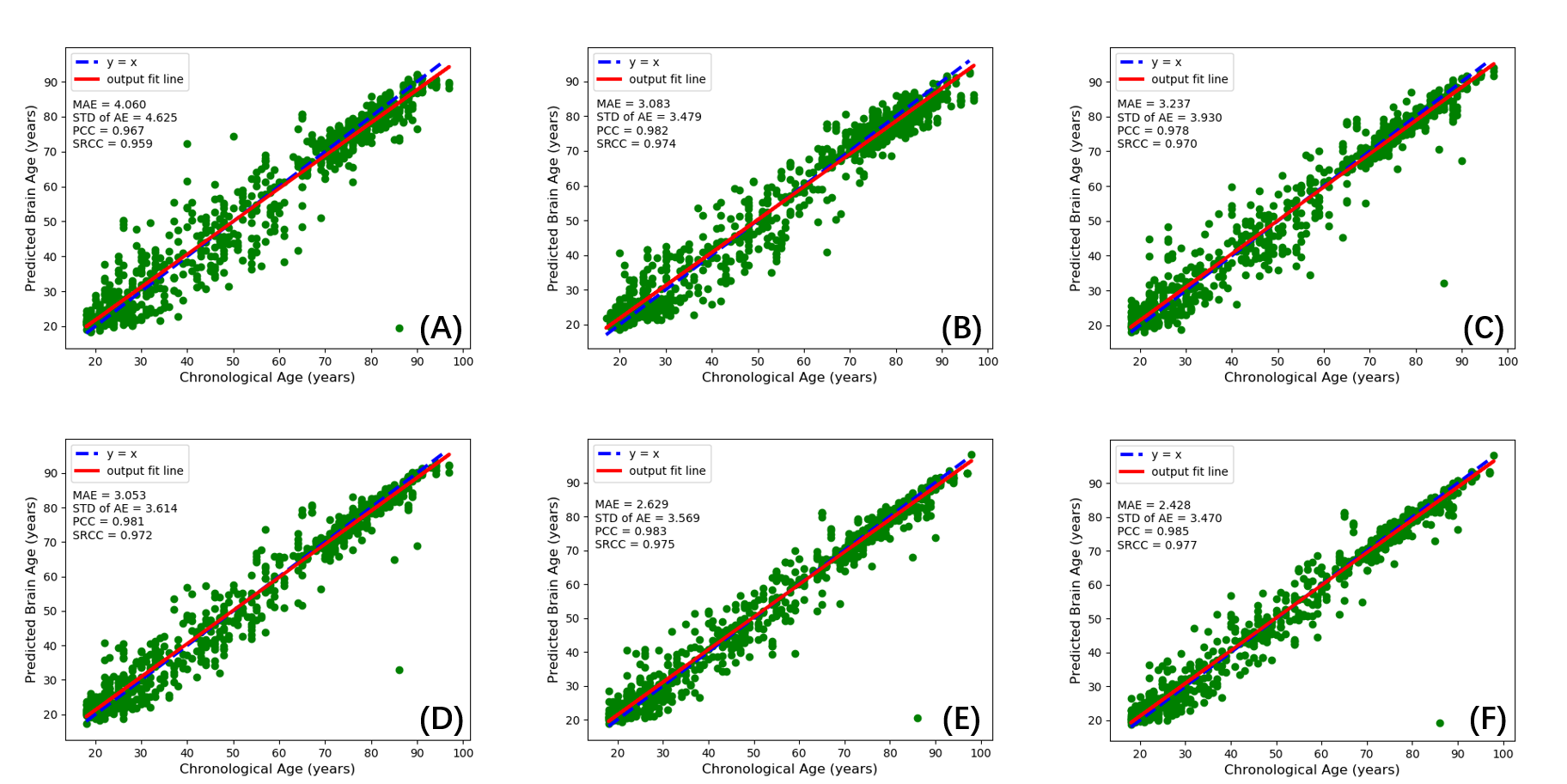}
  \caption{\label{fig:scatter}\textbf{Scatter diagrams of estimated brain ages in the test data by different estimation models.}  
  \textbf{(A)}: CNN in~\cite{cole:NI2017} with the MSE loss $\mathcal{L}_{\text{MSE}}$. 
  \textbf{(B)}: 
    The SFCN-transfer model in~\cite{Peng2019} with KL divergence and ranking losses.
  \textbf{(C)}: The first-stage network with AC and SE blocks, and the MSE loss $\mathcal{L}_{\text{MSE}}$. 
  \textbf{(D)}: The first-stage network with AC and SE blocks, and the total loss $\mathcal{L}$ in Eq.~\eqref{eq:loss}. 
  \textbf{(E)}: TSAN with the total loss $\mathcal{L}$ in Eq.~\eqref{eq:loss}. 
  \textbf{(F)}: Ensemble TSAN with the total loss $\mathcal{L}$ in Eq.~\eqref{eq:loss}. 
  In all diagrams, the blue dashed line indicates the ideal estimation $y=x$ (i.e., the estimated brain age equals the chronological age), 
  and the red line is the linear regression model fitted by the estimated brain age and chronological age.  
  MAE: mean absolute error; 
  STD of AE: the standard deviation of the absolute error; 
  PCC: PCC between the estimated brain age and chronological age, i.e., PCC$_\text{age}$; 
  SRCC: SRCC between the estimated brain age and chronological age, i.e., SRCC$_\text{age}$. 
  }
\end{figure*}

In Fig.~\ref{fig:scatter}, estimated brain ages in the test data by different brain age estimation models were plotted against chronological ages. 
For all models, a larger estimation error occurs in the chronological age range of $40$-$60$ years than other ranges, 
probably because we have less training data in this range, because of the unbalanced age distribution, as demonstrated in Fig.~\ref{fig:histogram_data}.  
STD of AE in Fig.~\ref{fig:scatter} shows the uncertainty of the estimation AE results.
The estimation results by different models were evaluated by MAE, PCC$_\text{age}$, SRCC$_\text{age}$, PCC$_\text{gap}$, and SRCC$_\text{gap}$ in Table~\ref{tab:bias}. 
Different loss functions, including the MAE loss $\mathcal{L}_{\text{MSE}}$, MSE loss $\mathcal{L}_{\text{MAE}}$, KL divergence, 
with or without ranking losses ($\mathcal{L}_{\text{d}}$ and $\mathcal{L}_{\text{r}}$), 
were also used to train CNN, SFCN, and TSAN. 
Table~\ref{tab:bias} demonstrates that compared with CNN in~\cite{cole:NI2017} and SFCN in~\cite{Peng2019}, the proposed TSAN generally obtains better results 
(i.e., lower MAE, higher PCC$_\text{age}$ and SRCC$_\text{age}$) in the brain age estimation task. 
  We mainly focus on MAE for evaluating brain age estimation, 
and ranking losses ($\mathcal{L}_d$ and $\mathcal{L}_r$) are used to regularize the network training and improve estimation MAE.
For all models (CNN, SFCN, and TSAN), the MSE loss generally obtains lower MAE than the MAE loss, and ranking losses help improving the MAE results. 
For all models, the estimation MAE is improved with ranking losses ($\mathcal{L}_d$ and $\mathcal{L}_r$) than without considering ranking losses.
Although CNN in~\cite{cole:NI2017} originally considered the MAE loss for training, the MSE loss obtains better result than the MAE loss, 
probability because MSE penalizes more for large errors when $\text{MAE}>1$, 
and MAE by existing methods is normally larger than $1$. 
For SFCN, the SFCN-pretrained model obtains worst MAE, because the testing data and training data are quite different. 
Moreover, SFCN-transfer yields best MAE among SFCN models, indicating that weight initialization for transfer learning and ranking losses improve the performance. 
For TSAN, compared with the traditional Conv block and MSE loss $\mathcal{L}_{\text{MSE}}$, 
the first-stage network with AC and SE blocks and the total loss $\mathcal{L}$ has better results. 
TSAN with two stage cascade networks obtains better estimation, compared with just using the first-stage network. 
The ensemble TSAN yields the best result with the lowest MAE ($2.428$), the largest PCC$_\text{age}$ and SRCC$_\text{age}$, 
which is significantly better than results by other models ($p<0.001$).

\subsection{Results with bias correction}

\begin{figure}[!t]
  \centering
  \includegraphics[width=\linewidth]{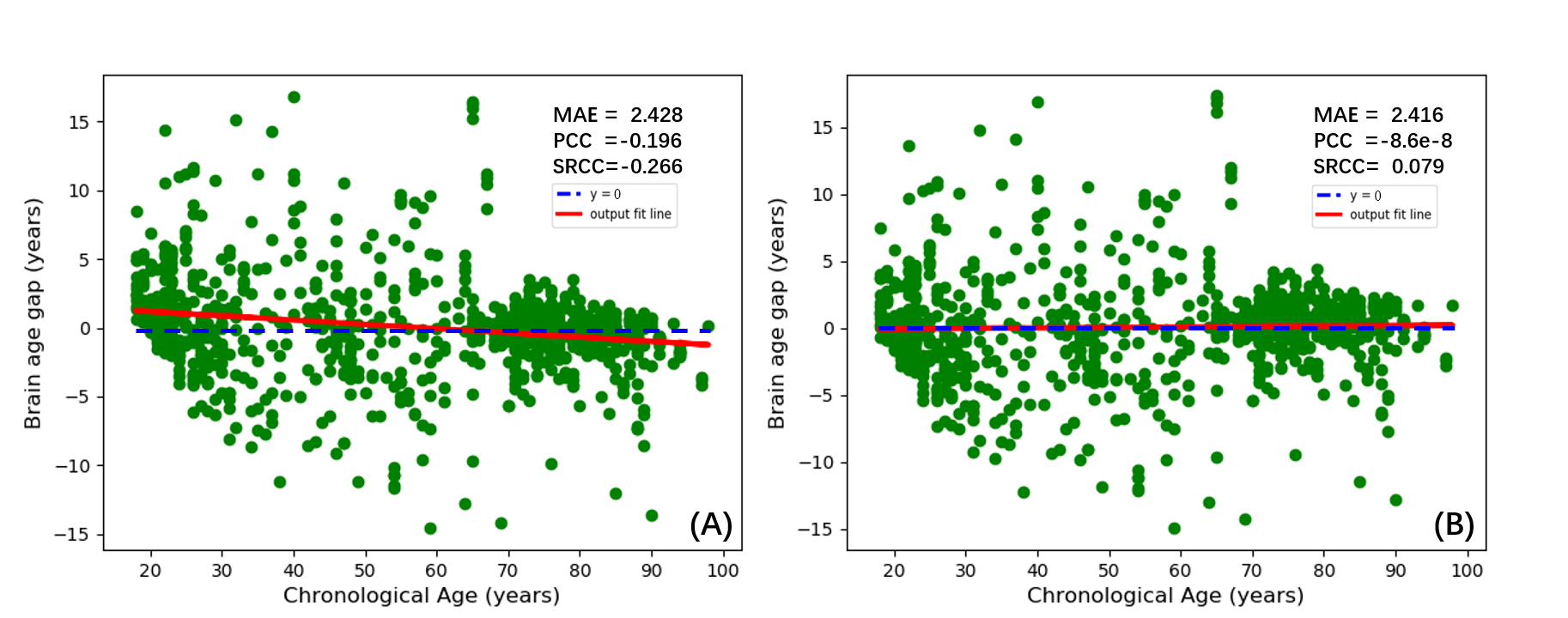}
  \caption{\label{fig:bias_correction}\textbf{Scatter plots of brain age gap versus chronological age on test data set, without and with linear bias correction.} 
  In both subfigures, the red line is the linear regression model fitted by the estimated brain age gap and chronological age
  and the dashed yellow line indicates the ideal estimation reference ($y=0$). 
  \textbf{(A)}: without bias correction. 
  \textbf{(B)}: with bias correction. 
  MAE: mean absolute error; 
  PCC: PCC between the estimated brain age gap and chronological age, i.e., PCC$_\text{gap}$;
  SRCC: SRCC between the estimated brain age gap and chronological age, i.e.,  SRCC$_\text{gap}$.}
\end{figure}

Fig.~\ref{fig:bias_correction} (A) shows the scatter plot of predicted brain age gap by the ensemble TSAN model versus chronological ages in the test data. 
There is a significant age-related variance in the brain age gap vs.\ chronological age without bias correction (PCC$_\text{gap}=-0.196$, SRCC$_\text{gap}=-0.266$, $p < 0.05$). 
We then performed bias correction by estimating a linear model in the training dataset and removing the linear bias for the test data as shown in Eq.~\eqref{eq:bias_correct}. 
Fig.~\ref{fig:bias_correction} (B) shows the scatter plot of the brain age gap vs.\ chronological age after applying bias correction by the proposed correction method.  
The correlation between brain age gap and chronological age after bias correction was close to zero (PCC$_\text{gap}=-8.6\times10^{-8}$, SRCC$_\text{gap}=-0.079$, $p < 0.05$), 
which indicates that the proposed bias correction method can successfully remove the age-related bias for estimation results.

We performed bias correction and calculated criteria for all models in Table~\ref{tab:bias}. 
After bias correction, PCC$_\text{gap}$ and SRCC$_\text{gap}$ are close to zero, as we expected. 
For all models, the estimated brain age with bias correction has consistently more accurate results (i.e., lower MAE, higher PCC$_\text{age}$ and SRCC$_\text{age}$), with lower bias (i.e., lower PCC$_\text{gap}$ and SRCC$_\text{gap}$).

\begin{table*}[!t]
  \caption{\label{tab:bias} Brain age estimation results in the test data by different estimation models with different losses, with and without bias correction, respectively. 
  }
  \centering
  \resizebox{0.85\width}{!}{%
  \begin{tabular}{c|c|ccccc|ccccc}
    \hline
    \multicolumn{1}{c|}{\multirow{3}{*}{Model}} & \multicolumn{1}{c|}{\multirow{3}{*}{Loss function}} & 
    \multicolumn{5}{|c}{Without Bias correction} & 
    \multicolumn{5}{|c}{With Bias correction}                                                                                                                                                                                                                      \\
    \cmidrule(r){3-7} \cmidrule(r){8-12}
    \multicolumn{1}{c|}{}& 
    \multicolumn{1}{c}{}& 
    \multicolumn{1}{|c}{MAE} & \multicolumn{1}{c}{\begin{tabular}[c]{@{}c@{}}PCC$_\text{age}$\end{tabular}} & \multicolumn{1}{c}{\begin{tabular}[c]{@{}c@{}}SRCC$_\text{age}$\end{tabular}} & 
      \multicolumn{1}{c}{\begin{tabular}[c]{@{}c@{}}PCC$_\text{gap}$\end{tabular}} & \multicolumn{1}{c}{\begin{tabular}[c]{@{}c@{}}SRCC$_\text{gap}$\end{tabular}} & 
        \multicolumn{1}{|c}{MAE} & \multicolumn{1}{c}{\begin{tabular}[c]{@{}c@{}}PCC$_\text{age}$\end{tabular}} & \multicolumn{1}{c}{\begin{tabular}[c]{@{}c@{}}SRCC$_\text{age}$\end{tabular}}&
          \multicolumn{1}{c}{\begin{tabular}[c]{@{}c@{}}PCC$_\text{gap}$\end{tabular}} & \multicolumn{1}{c}{\begin{tabular}[c]{@{}c@{}}SRCC$_\text{gap}$\end{tabular}}

     \\\hline CNN in~\cite{cole:NI2017}& MAE        & $4.455$ & $0.970$ & $0.954$ & $-0.312$ & $-0.383$ & $4.181$ & $0.975$ & $0.961$ & $-3.9\times10^{-8}$ & $-0.022$
            \\CNN in~\cite{cole:NI2017}& MSE        & $4.060$ & $0.967$ & $0.959$ & $-0.230$ & $-0.274$ & $3.969$ & $0.970$ & $0.965$ & $-2.7\times10^{-8}$ & $0.062$
            \\CNN in~\cite{cole:NI2017}& MAE + ranking       & $3.937$ & $0.968$ & $0.959$ & $-0.178$ & $-0.252$ & $3.881$ & $0.971$ & $0.962$ & $-1.0\times10^{-8}$ & $0.030$
            \\CNN in~\cite{cole:NI2017}& total loss (MSE + ranking)       & $3.133$ & $0.983$ & $0.975$ & $-0.227$ & $-0.279$ & $3.058$ & $0.984$ & $0.977$ & $-2.9\times10^{-8}$ & $0.017$
     \\\hline SFCN-pretrained~\cite{Peng2019}  &  KL divergence        & $4.927$ & $0.964$ & $0.938$ & $-0.301$ & $-0.376$ & $4.652$ & $0.969$ & $0.948$ & $3.8\times10^{-8}$  & $-0.034$
            \\SFCN-train~\cite{Peng2019}  &  KL divergence        & $3.941$ & $0.968$ & $0.959$ & $-0.186$ & $-0.261$ & $3.876$ & $0.971$ & $0.963$ & $1.4\times10^{-8}$  & $0.031$
            \\SFCN-transfer~\cite{Peng2019}  &  KL divergence        & $3.672$ & $0.979$ & $0.965$ & $-0.264$ & $-0.330$ & $3.583$ & $0.981$ & $0.969$ & $3.2\times10^{-8}$  & $-0.015$
            \\SFCN-train~\cite{Peng2019}  &  KL divergence + ranking        & $3.212$ & $0.982$ & $0.972$ & $-0.182$ & $-0.254$ & $3.146$ & $0.983$ & $0.974$ & $2.4\times10^{-8}$  & $0.002$
            \\SFCN-transfer~\cite{Peng2019}  &  KL divergence + ranking        & $3.083$ & $0.982$ & $0.974$ & $-0.202$ & $-0.252$ & $3.070$ & $0.983$ & $0.977$ & $2.2\times10^{-8}$  & $0.042$
     \\\hline First-stage network (Conv)  & MSE        & $3.666$ & $0.970$ & $0.964$ & $-0.197$ & $-0.256$ & $3.653$ & $0.972$ & $0.968$ & $2.9\times10^{-8}$  & $0.074$
            \\First-stage network  & MAE        & $4.899$ & $0.964$ & $0.938$ & $-0.294$ & $-0.369$ & $4.649$ & $0.969$ & $0.948$ & $-7.2\times10^{-8}$  & $-0.034$
            \\First-stage network  & MSE        & $3.237$ & $0.978$ & $0.970$ & $-0.204$ & $-0.243$ & $3.185$ & $0.979$ & $0.972$ & $\mathbf{-1.1\times10^{-8}}$  & $0.063$
            \\First-stage network  & MAE + ranking        & $3.080$ & $0.983$ & $0.975$ & $-0.202$ & $-0.245$ & $3.002$ & $0.984$ & $0.977$ & $2.0\times10^{-8}$  & $0.034$
            \\First-stage network  & total loss (MSE + ranking)        & $3.052$ & $0.981$ & $0.972$ & $\mathbf{-0.193}$ & $\mathbf{-0.241}$ & $3.010$ & $0.982$ & $0.974$ & $-4.4\times10^{-8}$  & $\mathbf{0.027}$
     \\\hline TSAN & MAE        & $2.940$ & $0.984$ & $0.975$ & $-0.210$ & $-0.267$ & $2.907$ & $0.985$ & $0.976$ & $4.8\times10^{-8}$  & $0.032$
            \\TSAN & MSE        & $2.937$ & $0.984$ & $0.975$ & $-0.202$ & $-0.253$ & $2.909$ & $0.985$ & $0.976$ & $4.3\times10^{-8}$ & $0.032$
            \\TSAN & MAE + ranking & $2.846$ & $0.985$ & $0.975$ & $-0.210$ & $-0.279$ & $2.806$ & $0.985$ & $0.978$ & $1.7\times10^{-8}$ & $0.031$
            \\TSAN & total loss (MSE + ranking) & $2.629$ & $0.983$ & $0.975$ & $-0.201$ & $-0.254$ & $2.617$ & $0.985$ & $0.976$ & $-3.3\times10^{-8}$ & $0.077$
            \\Ensemble TSAN & total loss (MSE + ranking) & $\mathbf{2.428}$ & $\mathbf{0.985}$ & $\mathbf{0.976}$ & $-0.196$ & $-0.266$ & $\mathbf{2.416}$ & $\mathbf{0.985}$ & $\mathbf{0.978}$ & $-8.6\times10^{-8}$ & $0.079$
            \\ \hline                               
  \end{tabular}
  }
\end{table*}

\begin{table*}[!t]
  \caption{\label{tab:sex} 
    Brain age estimation results of TSAN with and without sex labels.
  }
  \centering
  \resizebox{0.85\width}{!}{%
  \begin{tabular}{c|c|c|ccccc|ccccc}
    \hline
    \multicolumn{1}{c|}{\multirow{3}{*}{Model}} & \multicolumn{1}{c|}{\multirow{3}{*}{Loss function}} & \multicolumn{1}{c|}{\multirow{3}{*}{Sex}} &  
    \multicolumn{5}{|c}{Without Bias correction} & 
    \multicolumn{5}{|c}{With Bias correction}                                                                                                                                                                                                                      \\
    \cmidrule(r){4-8} \cmidrule(r){9-13}
    \multicolumn{1}{c|}{}& 
    \multicolumn{1}{c|}{}& 
    \multicolumn{1}{c|}{}& 
    \multicolumn{1}{|c}{MAE} & \multicolumn{1}{c}{\begin{tabular}[c]{@{}c@{}}PCC$_\text{age}$\end{tabular}} & \multicolumn{1}{c}{\begin{tabular}[c]{@{}c@{}}SRCC$_\text{age}$\end{tabular}} & 
      \multicolumn{1}{c}{\begin{tabular}[c]{@{}c@{}}PCC$_\text{gap}$\end{tabular}} & \multicolumn{1}{c}{\begin{tabular}[c]{@{}c@{}}SRCC$_\text{gap}$\end{tabular}} & 
        \multicolumn{1}{|c}{MAE} & \multicolumn{1}{c}{\begin{tabular}[c]{@{}c@{}}PCC$_\text{age}$\end{tabular}} & \multicolumn{1}{c}{\begin{tabular}[c]{@{}c@{}}SRCC$_\text{age}$\end{tabular}}&
          \multicolumn{1}{c}{\begin{tabular}[c]{@{}c@{}}PCC$_\text{gap}$\end{tabular}} & \multicolumn{1}{c}{\begin{tabular}[c]{@{}c@{}}SRCC$_\text{gap}$\end{tabular}}

            \\\hline First-stage network  & MSE  & No      & $3.394$ & $0.979$ & $0.969$ & $-0.226$ & $-0.266$ & $3.341$ & $0.981$ & $0.971$ & $\mathbf{-1.5\times10^{-8}}$  & $0.046$
            \\First-stage network  & MSE  & Yes      & $3.237$ & $0.978$ & $0.970$ & $-0.204$ & $-0.243$ & $3.185$ & $0.979$ & $0.972$ & $-2.7\times10^{-8}$  & $0.063$
            \\First-stage network  & total loss (MSE + ranking)  & No      & $3.129$ & $0.983$ & $0.975$ & $-0.232$ & $-0.287$ & $3.051$ & $0.985$ & $\mathbf{0.977}$ & $7.6\times10^{-8}$  & $\mathbf{0.016}$
            \\First-stage network  & total loss (MSE + ranking)  & Yes      & $3.052$ & $0.981$ & $0.972$ & $-0.193$ & $-0.241$ & $3.010$ & $0.982$ & $0.974$ & $-4.4\times10^{-8}$  & $0.027$
     \\\hline TSAN & MSE  & No      & $3.080$ & $0.983$ & $0.974$ & $-0.196$ & $\mathbf{-0.214}$ & $3.074$ & $0.984$ & $0.975$ & $2.9\times10^{-8}$  & $0.056$
            \\TSAN & MSE  & Yes      & $2.937$ & $\mathbf{0.984}$ & $0.975$ & $-0.202$ & $-0.253$ & $2.909$ & $0.985$ & $0.976$ & $4.3\times10^{-8}$ & $0.032$
            \\TSAN & total loss (MSE + ranking) & No & $2.918$ & $0.983$ & $0.975$ & $\mathbf{-0.192}$ & $-0.262$ & $2.883$ & $0.984$ & $0.976$ & $8.2\times10^{-8}$ & $0.030$
            \\TSAN & total loss (MSE + ranking) & Yes & $\mathbf{2.629}$ & $0.983$ & $\mathbf{0.975}$ & $-0.201$ & $-0.254$ & $\mathbf{2.617}$ & $\mathbf{0.985}$ & $0.976$ & $-3.3\times10^{-8}$ & $0.077$
            \\ \hline                               
  \end{tabular}
  }
\end{table*}

\begin{table*}[!t]
  \caption{\label{tab:delta} 
    Brain age estimation results of TSAN with different discretization parameter $\delta_d$.
  }
  \centering
  \resizebox{0.85\width}{!}{%
  \begin{tabular}{c|c|c|ccccc|ccccc}
    \hline
    \multicolumn{1}{c|}{\multirow{3}{*}{Model}} & \multicolumn{1}{c|}{\multirow{3}{*}{Loss function}} & \multicolumn{1}{c|}{\multirow{3}{*}{$\delta_d$}} &  
    \multicolumn{5}{|c}{Without Bias correction} & 
    \multicolumn{5}{|c}{With Bias correction}                                                                                                                                                                                                                      \\
    \cmidrule(r){4-8} \cmidrule(r){9-13}
    \multicolumn{1}{c|}{}& 
    \multicolumn{1}{c|}{}& 
    \multicolumn{1}{c|}{}& 
    \multicolumn{1}{|c}{MAE} & \multicolumn{1}{c}{\begin{tabular}[c]{@{}c@{}}PCC$_\text{age}$\end{tabular}} & \multicolumn{1}{c}{\begin{tabular}[c]{@{}c@{}}SRCC$_\text{age}$\end{tabular}} & 
      \multicolumn{1}{c}{\begin{tabular}[c]{@{}c@{}}PCC$_\text{gap}$\end{tabular}} & \multicolumn{1}{c}{\begin{tabular}[c]{@{}c@{}}SRCC$_\text{gap}$\end{tabular}} & 
        \multicolumn{1}{|c}{MAE} & \multicolumn{1}{c}{\begin{tabular}[c]{@{}c@{}}PCC$_\text{age}$\end{tabular}} & \multicolumn{1}{c}{\begin{tabular}[c]{@{}c@{}}SRCC$_\text{age}$\end{tabular}}&
          \multicolumn{1}{c}{\begin{tabular}[c]{@{}c@{}}PCC$_\text{gap}$\end{tabular}} & \multicolumn{1}{c}{\begin{tabular}[c]{@{}c@{}}SRCC$_\text{gap}$\end{tabular}}

    \\\hline  TSAN  & MSE  & 3      & $3.080$ & $0.983$ & $0.975$ & $-0.184$ & $-0.212$ & $3.054$ & $0.984$ & $0.977$ & $5.6\times10^{-8}$  & $0.032$
            \\TSAN  & MSE  & 5      & $2.937$ & $0.984$ & $0.975$ & $-0.202$ & $-0.253$ & $2.909$ & $0.985$ & $0.976$ & $4.3\times10^{-8}$  & $0.032$
            \\TSAN  & MSE  & 7      & $3.029$ & $0.984$ & $0.975$ & $-0.186$ & $-0.211$ & $2.994$ & $0.985$ & $0.977$ & $1.1\times10^{-8}$  & $0.045$
            \\TSAN  & MSE  & 9      & $2.940$ & $0.983$ & $0.975$ & $-0.211$ & $-0.265$ & $2.886$ & $0.985$ & $0.977$ & $3.9\times10^{-8}$  & $0.039$
     \\\hline TSAN & total loss (MSE + ranking) & 3 & $3.079$ & $0.982$ & $0.974$ & $-0.194$ & $-0.242$ & $3.071$ & $0.983$ & $0.976$ & $1.5\times10^{-8}$  & $0.042$
            \\TSAN & total loss (MSE + ranking) & 5 & $\mathbf{2.629}$ & $0.983$ & $0.975$ & $-0.201$ & $-0.254$ & $\mathbf{2.617}$ & $0.985$ & $0.976$ & $-3.3\times10^{-8}$ & $0.077$
            \\TSAN & total loss (MSE + ranking) & 7 & $2.764$ & $0.985$ & $0.978$ & $-0.197$ & $-0.245$ & $2.738$ & $0.986$ & $\mathbf{0.980}$ & $\mathbf{-1.0\times10^{-8}}$ & $0.048$
            \\TSAN & total loss (MSE + ranking) & 9 & $2.728$ & $\mathbf{0.986}$ & $\mathbf{0.978}$ & $\mathbf{-0.160}$ & $\mathbf{-0.209}$ & $2.731$ & $\mathbf{0.986}$ & $0.979$ & $-4.1\times10^{-8}$ & $\mathbf{0.015}$
            \\ \hline                               
  \end{tabular}
  }
\end{table*}

\begin{table*}[!t]
  \caption{\label{tab:layer} 
    Brain age estimation results of TSAN with different number of layers $N_{layer}$ and number of channels in the initial layer $N_{ini}$.
  }
  \centering
  \resizebox{0.85\width}{!}{%
  \begin{tabular}{c|c|c|ccccc|ccccc}
    \hline
    \multicolumn{1}{c|}{\multirow{3}{*}{Model}} & \multicolumn{1}{c|}{\multirow{3}{*}{Loss function}} & \multicolumn{1}{c|}{\multirow{3}{*}{$N_{layer}, N_{ini}$}} &  
    \multicolumn{5}{|c}{Without Bias correction} & 
    \multicolumn{5}{|c}{With Bias correction}                                                                                                                                                                                                                      \\
    \cmidrule(r){4-8} \cmidrule(r){9-13}
    \multicolumn{1}{c|}{}& 
    \multicolumn{1}{c|}{}& 
    \multicolumn{1}{c|}{}& 
    \multicolumn{1}{|c}{MAE} & \multicolumn{1}{c}{\begin{tabular}[c]{@{}c@{}}PCC$_\text{age}$\end{tabular}} & \multicolumn{1}{c}{\begin{tabular}[c]{@{}c@{}}SRCC$_\text{age}$\end{tabular}} & 
      \multicolumn{1}{c}{\begin{tabular}[c]{@{}c@{}}PCC$_\text{gap}$\end{tabular}} & \multicolumn{1}{c}{\begin{tabular}[c]{@{}c@{}}SRCC$_\text{gap}$\end{tabular}} & 
        \multicolumn{1}{|c}{MAE} & \multicolumn{1}{c}{\begin{tabular}[c]{@{}c@{}}PCC$_\text{age}$\end{tabular}} & \multicolumn{1}{c}{\begin{tabular}[c]{@{}c@{}}SRCC$_\text{age}$\end{tabular}}&
          \multicolumn{1}{c}{\begin{tabular}[c]{@{}c@{}}PCC$_\text{gap}$\end{tabular}} & \multicolumn{1}{c}{\begin{tabular}[c]{@{}c@{}}SRCC$_\text{gap}$\end{tabular}}

     \\\hline TSAN & total loss (MSE + ranking) & 4, 8  & $3.002$ & $0.983$ & $0.973$ & $-0.178$ & $-0.230$ & $2.989$ & $0.984$ & $0.975$ & $-2.6\times10^{-8}$  & $0.036$
            \\TSAN & total loss (MSE + ranking) & 5, 8  & $\mathbf{2.629}$ & $0.983$ & $0.975$ & $-0.201$ & $-0.254$ & $\mathbf{2.617}$ & $0.985$ & $0.976$ & $-3.3\times10^{-8}$ & $0.077$
            \\TSAN & total loss (MSE + ranking) & 6, 8  & $2.896$ & $0.984$ & $0.973$ & $\mathbf{-0.195}$ & $-0.248$ & $2.848$ & $0.985$ & $0.974$ & $\mathbf{1.0\times10^{-8}}$ & $0.025$
            \\TSAN & total loss (MSE + ranking) & 5, 4  & $3.168$ & $0.982$ & $0.972$ & $-0.238$ & $-0.311$ & $3.053$ & $0.983$ & $0.974$ & $8.9\times10^{-8}$ & $\mathbf{0.017}$
            \\TSAN & total loss (MSE + ranking) & 5, 16 & $2.661$ & $\mathbf{0.985}$ & $\mathbf{0.980}$ & $-0.197$ & $\mathbf{-0.242}$ & $2.659$ & $\mathbf{0.986}$ & $\mathbf{0.981}$ & $1.6\times10^{-8}$ & $0.065$
            \\ \hline                               
  \end{tabular}
  }
\end{table*}

\begin{figure}[!htbp]
  \centering
  \includegraphics[width=0.7\linewidth]{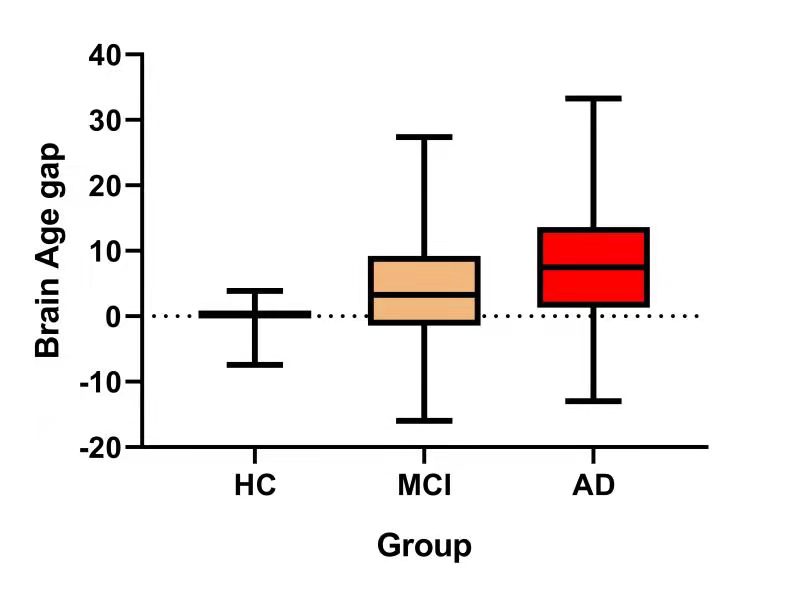}
  \caption{\label{fig:boxplot} \textbf{Comparison of brain age gap values of HC subjects, MCI patients (yellow box) and AD patients (red box).}  
  The mean brain age gap value for each group is illustrated with a solid black line in the middle of the box. 
  The reference dashed black line is $y=0$. 
  }
\end{figure}

\begin{table*}[!t]
  \caption{\label{tab:class_result_bias}Results of dementia classification based on brain age gap by different models, with and without bias correction, respectively. 
  }
  \centering
  \resizebox{0.7\width}{!}{%
  \begin{tabular}{c|c|c|cccc|cccc}
    \hline
    \multicolumn{1}{c|}{\multirow{3}{*}{Model}} & \multicolumn{1}{c|}{\multirow{3}{*}{Loss function}} & \multicolumn{1}{c|}{\multirow{3}{*}{Groups}} &  
    \multicolumn{4}{c|}{Without Bias correction} & 
    \multicolumn{4}{c}{With Bias correction}                                                                                                                                                                                                                      \\
    \cmidrule(r){4-7} \cmidrule(r){8-11}
    \multicolumn{1}{c|}{}& 
    \multicolumn{1}{c|}{}& 
    \multicolumn{1}{c|}{}& 
    \multicolumn{1}{|c}{AUC} & \multicolumn{1}{c}{\begin{tabular}[c]{@{}c@{}} ACC \end{tabular}} & \multicolumn{1}{c}{\begin{tabular}[c]{@{}c@{}} SEN \end{tabular}} & 
      \multicolumn{1}{c}{\begin{tabular}[c]{@{}c@{}} SPE \end{tabular}}  & 
    \multicolumn{1}{|c}{AUC} & \multicolumn{1}{c}{\begin{tabular}[c]{@{}c@{}} ACC \end{tabular}} & \multicolumn{1}{c}{\begin{tabular}[c]{@{}c@{}} SEN \end{tabular}} & 
      \multicolumn{1}{c}{\begin{tabular}[c]{@{}c@{}} SPE \end{tabular}}  \\ 

            \hline 
            CNN in~\cite{cole:NI2017} & MAE & HC vs.\ AD   & $0.743\pm 0.051$ & $0.600\pm 0.059$ & $0.525\pm 0.070$ & $0.963\pm 0.073$ & $0.765\pm 0.058$ & $0.652\pm 0.062$ & $0.593\pm 0.072$ & $0.938\pm 0.090$   \\
            CNN in~\cite{cole:NI2017} & MAE & HC vs.\ MCI  & $0.664\pm 0.063$ & $0.511\pm 0.050$ & $0.477\pm 0.061$ & $0.852\pm 0.143$ & $0.675\pm 0.057$ & $0.492\pm 0.046$ & $0.451\pm 0.052$ & $0.900\pm 0.115$   \\
            CNN in~\cite{cole:NI2017} & MAE & AD vs.\ MCI  & $0.556\pm 0.041$ & $0.610\pm 0.054$ & $0.716\pm 0.112$ & $0.395\pm 0.113$ & $0.563\pm 0.042$ & $0.613\pm 0.041$ & $0.710\pm 0.058$ & $0.418\pm 0.075$   \\
            \hline 
            CNN in~\cite{cole:NI2017} & MSE & HC vs.\ AD   & $0.741\pm 0.061$ & $0.638\pm 0.057$ & $0.582\pm 0.070$ & $0.901\pm 0.115$ & $0.747\pm 0.052$ & $0.614\pm 0.060$ & $0.544\pm 0.073$ & $0.952\pm 0.084$   \\
            CNN in~\cite{cole:NI2017} & MSE & HC vs.\ MCI  & $0.662\pm 0.056$ & $0.461\pm 0.048$ & $0.415\pm 0.054$ & $0.909\pm 0.113$ & $0.677\pm 0.058$ & $0.500\pm 0.049$ & $0.460\pm 0.058$ & $0.895\pm 0.125$   \\
            CNN in~\cite{cole:NI2017} & MSE & AD vs.\ MCI  & $0.571\pm 0.043$ & $0.594\pm 0.043$ & $0.638\pm 0.062$ & $0.503\pm 0.077$ & $0.578\pm 0.043$ & $0.574\pm 0.042$ & $0.568\pm 0.061$ & $0.588\pm 0.080$   \\
            \hline 
            CNN in~\cite{cole:NI2017} & total loss (MSE + ranking) & HC vs.\ AD   & $0.761\pm 0.062$ & $0.677\pm 0.066$ & $0.632\pm 0.083$ & $0.889\pm 0.120$ & $0.799\pm 0.056$ & $0.725\pm 0.057$ & $0.686\pm 0.073$ & $0.912\pm 0.113$   \\
            CNN in~\cite{cole:NI2017} & total loss (MSE + ranking) & HC vs.\ MCI  & $0.725\pm 0.055$ & $0.561\pm 0.048$ & $0.523\pm 0.055$ & $0.927\pm 0.114$ & $0.737\pm 0.055$ & $0.596\pm 0.046$ & $0.564\pm 0.054$ & $0.910\pm 0.113$   \\
            CNN in~\cite{cole:NI2017} & total loss (MSE + ranking) & AD vs.\ MCI  & $0.573\pm 0.043$ & $0.594\pm 0.041$ & $0.635\pm 0.061$ & $0.510\pm 0.083$ & $0.578\pm 0.041$ & $0.563\pm 0.039$ & $0.531\pm 0.059$ & $\mathbf{0.626\pm 0.081}$   \\
            \hline 
            SFCN-pretrained~\cite{Peng2019} & KL divergence & HC vs.\ AD   & $0.692\pm 0.057$ & $0.550\pm 0.063$ & $0.473\pm 0.079$ & $0.911\pm 0.107$ & $0.737\pm 0.064$ & $0.627\pm 0.062$ & $0.569\pm 0.072$ & $0.905\pm 0.108$   \\
            SFCN-pretrained~\cite{Peng2019} & KL divergence & HC vs.\ MCI  & $0.663\pm 0.058$ & $0.504\pm 0.050$ & $0.468\pm 0.057$ & $0.858\pm 0.120$ & $0.663\pm 0.055$ & $0.494\pm 0.048$ & $0.456\pm 0.055$ & $0.869\pm 0.108$   \\
            SFCN-pretrained~\cite{Peng2019} & KL divergence & AD vs.\ MCI  & $0.577\pm 0.038$ & $0.642\pm 0.042$ & $\mathbf{0.767\pm 0.075}$ & $0.387\pm 0.087$ & $0.596\pm 0.043$ & $0.661\pm 0.042$ & $\mathbf{0.786\pm 0.056}$ & $0.407\pm 0.074$   \\
            \hline 
            SFCN-transfer~\cite{Peng2019} & KL divergence + ranking & HC vs.\ AD   & $0.845\pm 0.036$ & $0.755\pm 0.052$ & $0.707\pm 0.063$ & $\mathbf{0.983\pm 0.043}$ & $0.868\pm 0.036$ & $0.785\pm 0.053$ & $0.741\pm 0.065$ & $\mathbf{0.994\pm 0.039}$   \\
            SFCN-transfer~\cite{Peng2019} & KL divergence + ranking & HC vs.\ MCI  & $0.730\pm 0.040$ & $0.538\pm 0.048$ & $0.495\pm 0.054$ & $\mathbf{0.966\pm 0.073}$ & $0.738\pm 0.041$ & $0.560\pm 0.048$ & $0.494\pm 0.054$ & $\mathbf{0.965\pm 0.072}$   \\
            SFCN-transfer~\cite{Peng2019} & KL divergence + ranking & AD vs.\ MCI  & $0.606\pm 0.041$ & $0.658\pm 0.759$ & $0.759\pm 0.055$ & $0.453\pm 0.075$ & $0.625\pm 0.042$ & $0.627\pm 0.041$ & $0.629\pm 0.055$ & $0.622\pm 0.074$   \\
            \hline 
            First-stage network & MSE & HC vs.\ AD    & $0.851\pm 0.051$ & $0.803\pm 0.051$ & $0.778\pm 0.063$ & $0.925\pm 0.094$ & $0.858\pm 0.048$ & $0.804\pm 0.056$ & $0.775\pm 0.070$ & $0.940\pm 0.090$   \\
            First-stage network & MSE & HC vs.\ MCI   & $0.773\pm 0.057$ & $0.706\pm 0.048$ & $0.691\pm 0.048$ & $0.855\pm 0.122$ & $0.786\pm 0.057$ & $0.706\pm 0.046$ & $0.703\pm 0.052$ & $0.869\pm 0.112$   \\
            First-stage network & MSE & AD vs.\ MCI   & $0.585\pm 0.042$ & $0.603\pm 0.045$ & $0.640\pm 0.083$ & $0.530\pm 0.082$ & $0.587\pm 0.042$ & $0.641\pm 0.044$ & $0.747\pm 0.069$ & $0.425\pm 0.082$   \\
            \hline 
            First-stage network & total loss (MSE + ranking) & HC vs.\ AD   & $0.877\pm 0.048$ & $0.834\pm 0.050$ & $0.811\pm 0.060$ & $0.943\pm 0.082$ & $0.893\pm 0.040$ & $0.861\pm 0.044$ & $0.844\pm 0.050$ & $0.942\pm 0.087$   \\
            First-stage network & total loss (MSE + ranking) & HC vs.\ MCI  & $0.796\pm 0.055$ & $0.749\pm 0.045$ & $0.735\pm 0.053$ & $0.878\pm 0.011$ & $0.806\pm 0.055$ & $0.749\pm 0.046$ & $0.736\pm 0.052$ & $0.878\pm 0.115$   \\
            First-stage network & total loss (MSE + ranking) & AD vs.\ MCI  & $0.634\pm 0.044$ & $0.653\pm 0.041$ & $0.690\pm 0.055$ & $\mathbf{0.577\pm 0.077}$ & $0.638\pm 0.044$ & $0.652\pm 0.042$ & $0.680\pm 0.062$ & $0.595\pm 0.083$   \\
            \hline 
            TSAN & total loss (MSE + ranking) & HC vs.\ AD   & $0.898\pm 0.044$ & $0.859\pm 0.044$ & $0.837\pm 0.058$ & $0.960\pm 0.072$ & $0.905\pm 0.045$ & $0.883\pm 0.040$ & $\mathbf{0.871\pm 0.040}$ & $0.940\pm 0.087$   \\
            TSAN & total loss (MSE + ranking) & HC vs.\ MCI  & $0.820\pm 0.054$ & $\mathbf{0.765\pm 0.052}$ & $\mathbf{0.752\pm 0.061}$ & $0.889\pm 0.116$ & $0.821\pm 0.051$ & $\mathbf{0.771\pm 0.046}$ & $\mathbf{0.759\pm 0.054}$ & $0.881\pm 0.112$   \\
            TSAN & total loss (MSE + ranking) & AD vs.\ MCI  & $0.618\pm 0.045$ & $0.642\pm 0.042$ & $0.688\pm 0.059$ & $0.548\pm 0.077$ & $0.625\pm 0.042$ & $0.630\pm 0.040$ & $0.639\pm 0.055$ & $0.612\pm 0.077$   \\
            \hline 
            Ensemble TSAN & Total loss & HC vs.\ AD   & $\mathbf{0.904\pm 0.041}$ & $\mathbf{0.868\pm 0.044}$ & $\mathbf{0.847\pm 0.054}$ & $0.962\pm 0.070$ & $\mathbf{0.919\pm 0.035}$ & $\mathbf{0.885\pm 0.042}$ & $0.864\pm 0.052$ & $0.974\pm 0.055$   \\
            Ensemble TSAN & Total loss & HC vs.\ MCI  & $\mathbf{0.823\pm 0.046}$ & $0.756\pm 0.044$ & $0.741\pm 0.052$ & $0.906\pm 0.098$ & $\mathbf{0.834\pm 0.046}$ & $0.753\pm 0.043$ & $0.735\pm 0.049$ & $0.934\pm 0.090$   \\
            Ensemble TSAN & Total loss & AD vs.\ MCI  & $\mathbf{0.639\pm 0.043}$ & $\mathbf{0.672\pm 0.043}$ & $0.734\pm 0.062$ & $0.545\pm 0.078$ & $\mathbf{0.646\pm 0.042}$ & $\mathbf{0.661\pm 0.041}$ & $0.693\pm 0.065$ & $0.598\pm 0.086$   \\
            \hline                               
  \end{tabular}
  }
\end{table*}

\subsection{Dementia classification}

Fig.~\ref{fig:boxplot} shows box plots of brain age gap values of all three groups (i.e., HC, AD, and MCI subject groups). 
The mean and standard deviation values of the brain age gap in the three groups are: 
$0.082 \pm 1.568$ for HC subjects, $3.517 \pm 7.694 $ for MCI patients, and $7.780 \pm 3.801$ for AD patients. 
The MAE and its standard deviation values in the three groups are: 
$1.067 \pm 1.150$ for HC subjects, $5.004 \pm 4.013 $ for MCI patients, and $9.471 \pm 4.124$ for AD patients. 
Both MCI and AD groups showed significantly higher brain age gap values compared with the HC group ($p < 0.001$). 

Table~\ref{tab:class_result_bias} shows the results of the dementia classification experiments by using brain age gap from different models as the only input variable of SVM, with and without bias correction, respectively. 
The mean and STD results were calculated based on results of nested CV for $3000$ times. 
It can been seen from Table~\ref{tab:class_result_bias} that the proposed ensemble TSAN with the total loss and bias correction generally yields the best performance. 
For HC vs.\ AD, ensemble TSAN has the best performance with AUC of $0.904 \pm 0.041$, accuracy of $86.8\%\pm 4.4\%$, sensitivity of $84.7\%\pm 5.4\%$, and specificity of $96.2\%\pm 7\%$. 

\section{Discussion}

Compared with existing brain age estimation methods~\cite{Peng2019,Gaser2013covert,Li2018,Herent2018,franke:NI2012:age,Smith2019,franke2012longitudinal,cole:NI2017,Lin2016,Cole2017a,franke:NI2010:age}, 
the proposed TSAN is a novel 3D convolutional neural network with a two-stage cascade architecture for brain age estimation from MRIs. 
See the architecture of TSAN in Fig.~\ref{fig:TSAN} and Fig.~\ref{fig:TSAN_network}. 
Note that due to the discretization in the first-stage net, the first-stage network only needs a rough estimation. 
Then, the second-stage net actually estimates a residual age which is then added to the discretized output brain age by the first-stage net. 
In this way, we found it obtains better experimental results than the traditional cascade network architecture without discretization and the residual path. 

Two ranking losses were proposed for regularizing the training of TSAN, besides the traditional MSE loss. 
The MSE loss is defined using the estimated brain age and true age from only an individual sample. 
While the age difference loss (i.e., the first ranking loss) is defined based on a pair of samples, 
and the second ranking loss is based on SRCC of a set of samples. 
In this way, the two ranking losses could consider coupled relationship among all samples in a training batch, 
which yields better experimental performances as shown in Table~\ref{tab:bias} and Table~\ref{tab:class_result_bias}.  
The proposed two ranking losses improve not only the proposed TSAN, but also CNN in~\cite{cole:NI2017} and SFCN in~\cite{Peng2019}.

A larger training dataset and a smaller chronological age range in the test data normally result in a better estimation with smaller MAE~\cite{cole:NI2020}. 
Peng \emph{et al.}~\cite{Peng2019} proposed Simple Fully Convolutional Network (SFCN) and won Predictive Analytics Competition 2019 (PAC-2019) with $\text{MAE}=2.90$. 
The ensemble TSAN obtained $\text{MAE}=2.428$ for the combined data with $6586$ MRIs, which is lower than SFCN for PAC-2019 data with $\text{MAE}=2.90$~\cite{Peng2019}, 
although the test data of PAC-2019 was not used for evaluating TSAN, because the ground true age labels were not released by organizers. 
SFCN obtained $\text{MAE}=2.14$ in UK Biobank data~\cite{Peng2019}, 
while UK Biobank data has a larger number of MRIs $N=14503$ and smaller age range of $44$ - $80$, 
compared with the combined dataset with $6586$ MRIs and age range of $17$ - $98$ used in this paper. 
Moreover, UK Biobank dataset is a very homogeneous dataset from the same type of scanner with the same scanning protocol. 
Thus, brain age estimation in UK Biobank data may be easier than other datasets like the combined dataset with ADNI-1, PAC-2019 and OASIS.
Applying the proposed TSAN and ranking losses in the UK Biobank data is beyond the scope of this paper, and could be considered in our future work.

Experimentally we found that the estimated brain age is greater than the chronological age for youthful subjects and is smaller for older subjects. 
See Fig.~\ref{fig:bias_correction} (A). 
The age-related bias occurs probably because of inconsistency of noise distribution across the lifespan, 
sample size imbalance across age groups, or heterogeneity of data from a couple of learn about sites. 
Note that the bias does not occur only for specific methods. 
Liang \emph{et al.}~\cite{Liang2019} tested four popular machine learning techniques including ridge regression, 
support vector regression, Gaussian processes regression and deep neural network, and they all showed a systematic bias in brain age estimation.
Le \emph{et al.}~\cite{le2018nonlinear} proposed a nonlinear simulation framework to reduce the age-related bias. 
We applied bias correction to brain age estimation based on linear regression model, 
which can reduce the variance in brain age gap and lead to lower MAE after correction. 
See Fig.~\ref{fig:bias_correction} (B). 

As shown in Fig.~\ref{fig:boxplot}, brain age gap estimated by T1-weighted MRI data demonstrated its relationship associated with brain aging. 
With the disease progresses, brain age is statistically larger than chronological age: 
the mean brain age gap in MCI patients was about 3 years greater than in healthy samples, 
while it was about 7 years higher in AD patients than in healthy samples. 
This allows us to efficiently distinguish AD or MCI patients from healthy subjects.
Finally, brain age gap was used as the only input variable of SVM to classify healthy control subjects, AD and MCI patients. 
Dementia classification is a promising application of brain age estimation. 
Table~\ref{tab:bias} and Table~\ref{tab:class_result_bias} demonstrate that the proposed ensemble TSAN with the total loss and bias correction 
obtains the best result in both age estimation (with lowest MAE, highest PCC$_\text{age}$, and highest SRCC$_\text{age}$) 
and dementia classification (with highest AUC, ACC, SEN and SPE).

\section{Conclusion}

In this paper, a novel 3D convolutional network, called two-stage-age-network (TSAN), is proposed to estimate brain age from T1-weighted MRIs. 
TSAN uses a two-stage cascade architecture, 
where the second-stage network is to refine the estimated age based on the discretized output of the first-stage network. 
In both stages, we propose a ScaledDense block to concatenate feature maps from different scales,  
and consider both MR images and sex labels as inputs of the networks. 
To our knowledge, TSAN is the first work to apply novel ranking losses in brain age estimation, together with the traditional MSE loss. 
One ranking loss is based on age difference of paired samples, and the other one is calculated based on Spearman's rank correlation coefficient (SRCC) from a batch of training samples. 
Furthermore, we perform bias correction to brain age by the linear regression, 
which provides more accurate estimation of brain age and brain age gap. 
The proposed TSAN was validated using $6586$ MRIs from three datasets, 
yielding an MAE of 
$2.428$, PCC of $0.985$, and SRCC of $0.976$, 
between the estimated brain age and the chronological ages. 
At last, we validated brain age estimation by using brain age gap as the only input variable of SVM to classify healthy control subjects, MCI and AD patients. 
The proposed ensemble TSAN after bias correction generally yields the best classification performance in terms of AUC, accuracy, sensitivity and specificity. 
It confirms that brain age is a promising biomarker for dementia classification or early-stage dementia risk screening.


\end{document}